\documentclass[aps,pra,reprint,superscriptaddress,letterpaper,twocolumn]{revtex4-1}
\usepackage[english]{babel}
\usepackage{ucs}
\usepackage[utf8x]{inputenc}
\usepackage{amsmath}
\usepackage{amssymb}
\usepackage{amsthm}
\usepackage{mathrsfs}
\usepackage{graphicx}
\usepackage{bm}
\usepackage{bbm}
\usepackage{empheq}
\usepackage{cases}
\usepackage{euscript}
\usepackage[usenames, dvipsnames]{color}
\usepackage[colorlinks=true,linkcolor=blue,citecolor=blue,urlcolor=blue]{hyperref}
\usepackage{enumitem}

\newcommand{\bra}[1]{\langle #1|}
\newcommand{\ket}[1]{|#1\rangle}

\newcommand{\ketbra}[2]{\ket{#1}\!\bra{#2}}
\newcommand{\norm}[1]{\left\lVert#1\right\rVert}
\newcommand{\mm}[1]{\mathrm{#1}}
\newcommand{\ui}{\mathrm{i}}
\newcommand{\ub}{\mathrm{b}}
\newcommand{\up}{\mathrm{p}}
\newcommand{\us}{\mathrm{s}}
\newcommand{\ue}{\mathrm{e}}
\newcommand{\uf}{\mathrm{f}}
\newcommand{\ud}{\mathrm{d}}

\newcommand{\uI}{\mathrm{I}}
\newcommand{\uG}{\mathrm{G}}

\newcommand{\rd}{\partial}

\newcommand{\abs}[1]{\left|#1\right|}

\newcommand{\com}[2]{\left[#1,#2\right]}

\def \hH{\widehat{H}}
\def \hXi{\widehat{\Xi}}
\def \hzeta{\widehat{\zeta}}
\def \hV{\widehat{V}}
\def \hU{\widehat{U}}
\def \hO{\widehat{O}}
\def \hT{\widehat{T}}
\def \hS{\widehat{S}}
\def \hR{\widehat{R}}
\def \hW{\widehat{W}}
\def \hX{\widehat{X}}
\def \hP{\widehat{P}}
\def \hOmega{\widehat{\Omega}}
\def \pp{\epsilon}
\def \HHcomp{\mathcal{H}_{\rm comp}}
\def \hZ{\widehat{Z}}
\def \sq{\mathcal{Q}}

\def \sL{\EuScript{L}}
\DeclareFontFamily{OT1}{pzc}{}
\DeclareFontShape{OT1}{pzc}{m}{it}{<-> s * [1.10] pzcmi7t}{}
\DeclareMathAlphabet{\mathpzc}{OT1}{pzc}{m}{it}

\begin{document}

\title{Systematic Magnus-based approach for suppressing leakage and non-adiabatic errors in quantum
dynamics}

\author{Hugo Ribeiro}
\affiliation{Department of Physics, McGill University, 3600 rue University, Montréal, Qc H3A 2T8, Canada}

\author{Alexandre Baksic}
\affiliation{Department of Physics, McGill University, 3600 rue University, Montréal, Qc
H3A 2T8, Canada}

\author{Aashish A. Clerk}
\affiliation{Department of Physics, McGill University, 3600 rue University, Montréal,
Qc H3A 2T8, Canada}

\begin{abstract}
We present a systematic, perturbative method for correcting quantum gates to suppress errors that
take the target system out of a chosen subspace. It addresses the generic problem of non-adiabatic
errors in adiabatic evolution and state preparation, as well as general leakage errors due to spurious
couplings to undesirable states. The method is based on the Magnus expansion: by correcting control
pulses, we modify the Magnus expansion of an initially-given, imperfect unitary in such a way that
the desired evolution is obtained.  Applications to adiabatic quantum state transfer,
superconducting qubits and generalized Landau-Zener problems are discussed. 
\end{abstract}

\maketitle

\section{Introduction}
\label{sec:intro}

The problem of leakage errors, where a quantum gate is corrupted by populating spurious states,
is generic to a variety of situations in quantum information processing. The most prominent example
is the problem of high-fidelity qubit gates.  Most physical implementations of qubits are
multi-level systems, with two levels chosen to encode the logical states of the qubit.  As a result,
control sequences designed to implement a given unitary evolution can give rise to transitions out of the
logical subspace (see, e.g.~\cite{chen2016}).  Such ``leakage errors'' generically become more
prominent as gates are made faster, due to the increased bandwidth of control pulses (and consequent
enhanced spectral weight at the frequencies of unwanted transitions).  Yet another generic example
of leakage errors comes from protocols utilizing adiabatic evolution, where e.g. one attempts to
have a system remain in the instantaneous ground state of some time dependent Hamiltonian.  The
leakage errors here occur for any non-zero protocol speed, and take the form of non-adiabatic
transitions between instantaneous eigenstates.

Given their ubiquity, there is great interest in devising methods for suppressing leakage errors.
While one could use the machinery of optimal quantum control~\cite{khaneja2005}, analytic approaches
which yield simple, smooth and robust control pulses are also desirable. Many approaches have been
put forth in the context of specific problems. For superconducting circuits, perhaps the best known
is the DRAG technique (Derivative Removal by Adiabatic Gate)~\cite{motzoi2009,gambetta2011}, which
was designed to minimize leakage errors for single qubit gates in a weakly anharmonic qubit. DRAG
has also been employed to improve the fidelity of a Rydberg-blockade two-qubit entangling
gate~\cite{theis2016}. A technique for minimizing leakage errors in two-qubit gates in circuit QED
has also been formulated \cite{economou2015}.  Turning to leakage errors in adiabatic-evolution
protocols, a general strategy here are the so-called ``shortcuts to adiabaticity'' (STA) techniques
\cite{demirplak2003,demirplak2008,berry2009,ibanez2012,chen2012,torrontegui2013,baksic2016}, which
provide methods for improving pulse sequences. A recent technique based on a ``quantum geometric''
interpretation of adiabatic evolution has also been put forward~\cite{tomka2016}. 

In this work we present an extremely general strategy for mitigating leakage errors.  In
certain limits, it captures aspects of both DRAG and STA approaches, but is also able to deal
with situations where these methods fail or become impossible to implement. 

\begin{figure}[t!]
	\includegraphics[width=\columnwidth]{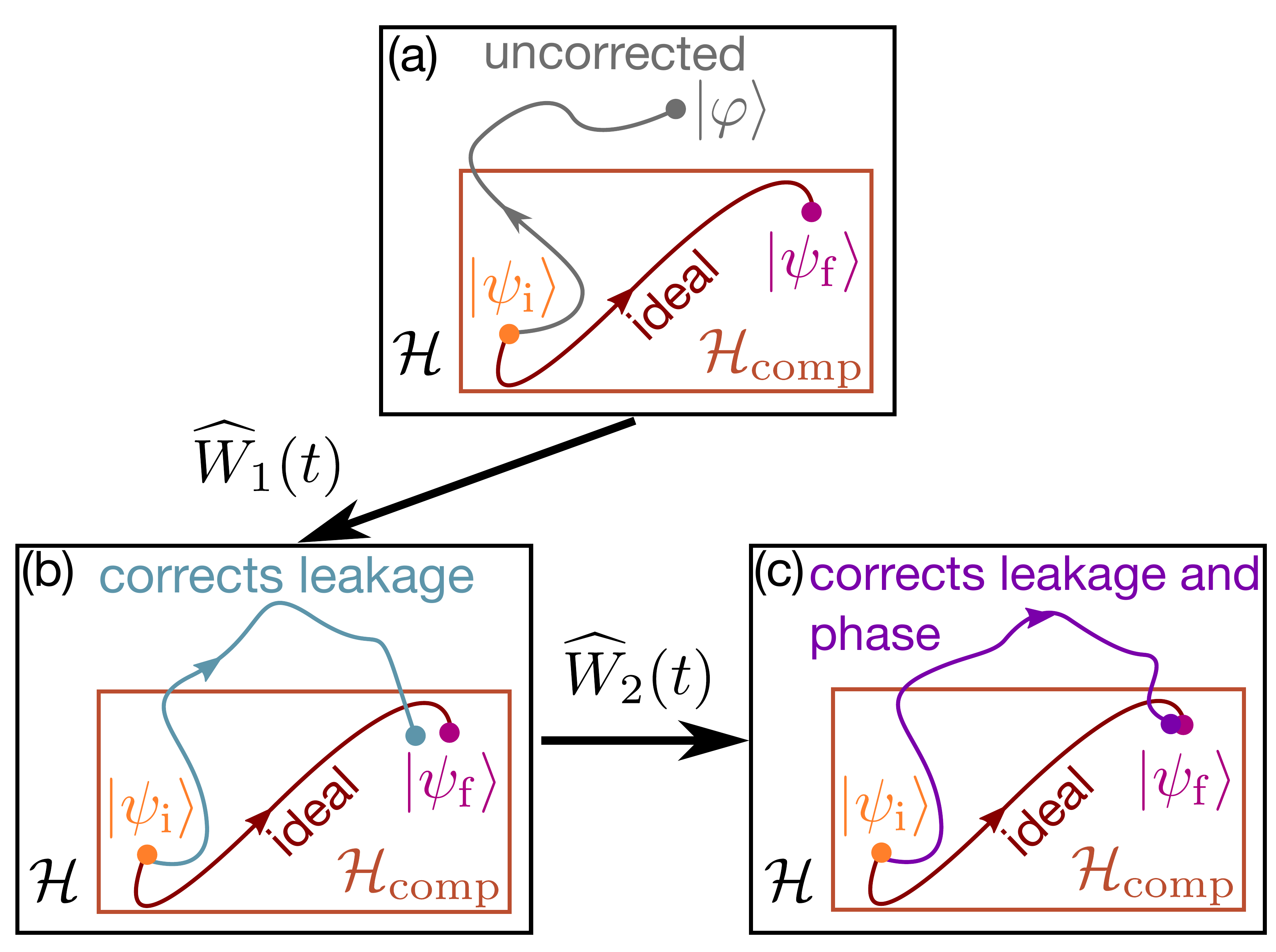}
	\caption{(Color online). Schematic representation of the Magnus-based algorithm. (a) Before
		applying the algorithm, the ideal evolution is corrupted by the spurious coupling,
		which prevents to reach the target state $\ket{\psi_\uf}$. (b) After the first step
		of the Magnus-based algorithm, ``pure'' leakage errors are cancelled on average by
		the control Hamiltonian $\hW_1 (t)$. Although the system ends in
		$\mathcal{H}_{\mm{comp}}$ at $t_\uf$, the target state $\ket{\psi_\uf}$ is not
		reached. (c) The second step of the Magnus-based algorithm corrects induced phase
		errors within $\mathcal{H}_{\mm{comp}}$ with the help of the control Hamiltonian $\hW_2
		(t)$.}
	\label{fig:idea} 
\end{figure}

The underlying idea is sketched schematically in Fig.~\ref{fig:idea}. We start with a control
sequence that would perfectly implement some desired quantum gate if there were no leakage
levels. This evolution is however corrupted by coupling to spurious leakage levels
[Fig.~\ref{fig:idea}~(a)]. Our goal is to gently modify the control sequence to mitigate the errors.
We do this in two steps. First, we design a control correction [described by a Hamiltonian
$\hW_1(t)$] that cancels the effects of pure leakage to leading order.  With this correction, the
population of leakage levels at the final time is highly suppressed.  The dominant remaining error
is then a ``phase" error:  despite residing in the computational subspace, the final state may still
deviate from the target state [Fig.~\ref{fig:idea}~(b)]. Our second order control correction
[described by $\hW_2(t)$] corrects for these errors [Fig.~\ref{fig:idea}~(c)]. We stress that the
goal is not to suppress leakage errors at all times, but rather to have them effectively cancel out
{\it on average.}

The general philosophy sketched above is in many ways similar to that of both DRAG and  STA (see,
e.g., Refs.~\cite{motzoi2009,baksic2016}).  However, our means of implementing it is quite
different.  We use the Magnus expansion~\cite{magnus1954,blanes2009} to systematically improve
control pulses and implement our strategy of cancelling leakage on average. As we show in detail,
this allows us to treat a more general class of problems. In particular, we are able to correct
leakage errors in cases where the DRAG approach fails due to the closing of a spectral gap. Also, in
comparison to STA techniques, we can apply our methodology in a perturbative fashion, allowing us to
treat complex systems where the exact diagonalization usually employed in STA (e.g. to find
counterdiabatic driving
fields~\cite{demirplak2003,demirplak2008,berry2009,ibanez2012,chen2012,baksic2016}) is impractical.
Recently a numerical variational approach has been put forward to circumvent the need of exact
diagonalization~\cite{sels2016}.

The general idea of using the Magnus expansion to find optimized control sequences is of course not
new, with the earliest work addressing the specific problem of achieving population inversion in
nuclear magnetic resonance~\cite{tycko1983}. Most Magnus-based approaches for improved controls rely
on simply using the first term of the expansion, and viewing this as a Fourier transform integral;
one then corrects pulses to suppress spectral weight associated with unwanted leakage
transitions~\cite{warren1984,schutjens2013}. Similar approaches based on Fourier analysis have been
explored in the context of specific problems in Refs.~\cite{hansel2001} and \cite{martinis2014}. As
we discuss, our Magnus approach is significantly different from these strategies:~it does not rely
on an analogy to a Fourier transform (allowing us to treat more general problems), and is not
restricted to simply looking at the first order term in the Magnus expansion.     

The remainder of this paper is organized as follows: in Sec.~\ref{sec:generalproblem} we define the
general ``leakage problem" in the context of a quantum gate generated by a Hamiltonian with
time-dependent control fields. In Sec.~\ref{sec:magnussol}, we show how to systematically correct
the control fields using the Magnus expansion so as to mitigate leakage errors.  Finally, in
Sec.~\ref{sec:applications}, we consider a few illustrative and relevant applications of the
method:~suppression of non-adiabatic errors in stimulated Raman adiabatic passage (STIRAP) in a
lambda-system, suppression of leakage in gates for superconducting qubits, and suppression of
leakage in the multiple-crossings Landau-Zener model. 

Our choice of examples in Sec.~\ref{sec:applications} has been made to clearly contrast the
Magnus-based approach against DRAG and STA techniques. The relatively simple problem of STIRAP
demonstrates how the Magnus approach lets one perturbatively capture the leading effect of the STA
technique; it also illustrates how DRAG fails when confronted with the closing of a spectral gap.
Re-visiting the problem of leakage errors in superconducting qubits (for which DRAG was developed)
helps one interpret the recent experimental result of Chen \textit{et al.}~\cite{chen2016}.  This
work demonstrated that optimizing the overall amplitudes of the DRAG controls (away from the
theoretically-predicted optimal values) helps further reduce gate errors; our Magnus-based approach
provides a natural way for understanding why this is the case. Finally, the multiple-crossings
Landau-Zener model demonstrates how the Magnus approach can correct a complex adiabatic evolution
problem for which STA techniques cannot be implemented.  

\section{Statement of the general leakage error problem}
\label{sec:generalproblem}

We consider a quantum system with a Hilbert space $\mathcal{H}$, having a subspace $\HHcomp$ that
contains the useful computational states of the system.  The goal is to implement some specified
unitary gate $\hU_\uG$ taking $\HHcomp \rightarrow \HHcomp$.  The basic problem is that the
time-dependent controls used to generate this evolution will invariably couple states inside
$\HHcomp$ to those outside of it, leading to leakage and an imperfect implementation of the quantum
gate.  

For concreteness, we consider a basis $\{ \ket{n}, 1 \leq n \leq N \}$, where the first $Q$ levels
form the computational subspace  (e.g.~the coupled states of a qubit encoded in $\ket{0}$ and
$\ket{1}$), and the remaining $N-Q$ levels are ``leakage'' levels.  The generic time-dependent
Hamiltonian will have the form
\begin{equation}
	\begin{aligned}
		\hH(t) &= \hH_0 (t) + \epsilon \hV(t) \\
		\hH_0 (t) & = \hH_{Q}(t) + \sum_{n=Q+1}^N E_n(t) \ketbra{n}{n}.
	\end{aligned}
	\label{eq:Hgen}
\end{equation}
$\hH_{Q}(t)$ represents the desired control of the computational levels, and only has non-zero
matrix elements between these levels.  $\hV(t)$ describes both the spurious couplings that are
generated between computational levels and leakage levels (e.g. for a qubit, the coupling between
$\ket{0}$ or $\ket{1}$ with any other excited state $\ket{n}$, $n\geq2$) and couplings between
leakage states \footnote{Note that we keep the couplings between leakage states so that the
interaction picture generated by \unexpanded{$\hH_0(t)$} is as simple as possible.  While
keeping these couplings in \unexpanded{$\hH_0$} might appear natural, it would in general
lead to an interaction picture that is too complicated to easily work with in all but the
simplest problems}.  We include the dimensionless parameter $\pp$ here for convenience; it will let
us track powers of $\hV(t)$ in what follows. 

Solving the Schrödinger equation defined by Eq.~\eqref{eq:Hgen} leads to the unitary time evolution
operator 
\begin{equation}
	\hU(t) = \hU_0 (t) \hU_\uI (t).
	\label{eq:Ugen}
\end{equation}
Here, $\hU_0(t)$ describes the desired evolution generated by $\hH_0 (t)$,
\begin{equation}
	\hU_0 (t)=\hT \exp\left[-i \int_0^t \ud t_1 \hH_0 (t_1)\right],
	\label{eq:U0}
\end{equation}
with $\hT$ the time-ordering operator. For simplicity and without loss of generality, we have chosen
to set $t_\ui = 0$. In contrast, imperfections due to leakage are described by $\hU_\uI (t)$, which
is defined as
\begin{equation}
	\hU_\uI (t) = \hT\exp\left[-i  \epsilon \int_0^t \ud t_1 \hV_\uI (t_1) \right].
	\label{eq:UI}
\end{equation}
\begin{center}
\begin{table*}
	\caption{Definition of the most important symbols}
\begin{tabular}{ c c c }
	  \hline
	  Symbol & Meaning  & Equation \\
	  \hline
	  \hline
	  $\hV (t)$  & error Hamiltonian & Eq.~\eqref{eq:Hgen}\\
	  $\hO_\uI(t)$ & operator in the interaction picture generated by $\hH_0(t)$ & Eq.~\eqref{eq:SuperOpL} \\
	  $\hW_j (t)$ & Hamiltonian describing $j$th-order correction to control the fields  & Eq.~\eqref{eq:WSeries} \\
	  $\hV^{(j)} (t) $ & modified error Hamiltonian (includes first $j$ control Hamiltonians  $\hW_j (t)$) &
	  Eq.~\eqref{eq:moderrprop} \\
  	  $\hOmega^{(j)} (t)$ & modified ``Magnus operator'' associated to $\hV_\uI^{(j)} (t)$ &
   	  Eq.~\eqref{eq:moderrpropMagnus} \\
	  $\hOmega^{(j)}_k (t)$ & $k$th term of the modified Magnus expansion &
	  Eq.~\eqref{eq:magnus_series}
\end{tabular}
\end{table*}
\end{center}
Here and throughout, we denote the interaction picture representation of a Schrödinger operator
$\hO(t)$ by $\hO_\uI (t)$:
\begin{equation}
	\hO_\uI(t) = \hU_0^\dag (t) \hO(t) \hU_0(t) = \ell_0 (t) \hO(t).
	\label{eq:SuperOpL}
\end{equation}	
The last inequality defines the superoperator $\ell_0(t)$.  Note that this interaction picture is in
general easy to find: one only needs to fully understand the desired dynamics in the computational
subspace, as the evolution of the leakage subspace is trivial under $\hH_0(t)$\footnote{We note that
it is not necessary to perform the interaction-picture with respect to the evolution defined
by \unexpanded{$\hH_0 (t)$}.  One can also perform the more conventional interaction-picture
transformation with \unexpanded{$\hH_0 (t)$} a diagonal operator. In this case, we have
\unexpanded{$\hH (t) = \hH_0 (t) + \hV_0 (t) + \hV (t)$ with $\hV_0 (t)$} the interaction
allowing to manipulate the states within $\HHcomp$. The controls must then be chosen such
that \unexpanded{$\hOmega^{(j)} (t_\uf) = \hOmega^{(j)} [\hV_0 (t_\uf)] +
\mathcal{O}(\epsilon^{j+1})$}, i.e. we do not cancel the controls necessary to perform
the unitary gate operation.}. We also emphasize that the interaction picture defined in
Eq.~\eqref{eq:SuperOpL} is different than the one used to derive
DRAG~\cite{motzoi2009,gambetta2011}. Our interaction picture is defined by the full unitary
evolution generated by the ideal Hamiltonian. As a consequence, our approach treats only the
``spurious'' couplings as a perturbation (and not the desired couplings between computational
levels).

We assume that the various time-dependent fields have been chosen such that if there were no
leakage, the evolution described by $\hU_0 (t)$ at $t = t_\uf$ corresponds to some desired unitary
gate $\hU_\uG$ in the computational subspace.  Formally, in the case where there was no
coupling to the leakage subspace, we would have:
\begin{equation}
	 \hP_Q \hU_\uI (t) \hP_Q = \mathbbm{1}_Q,
\end{equation}
and achieving the desired gate would require
\begin{equation}
	   \hP_Q \hU(t_\uf) \hP_Q =   \hP_Q \hU_0(t_\uf) \hP_Q = \hU_\uG.
	   \label{eq:PerfectGate}
\end{equation}
We have introduced here the projection operator onto the computational subspace, $\hP_Q$, and unit
operator in this subspace $\mathbbm{1}_Q$.

Including the effects of non-zero leakage, the ideal gate will be corrupted, and states initially in
the computational subspace will evolve to states outside of it [see Fig.~\ref{fig:idea}~(a)].  To
mitigate the effects of leakage, one could try to somehow suppress the relevant off-diagonal
elements of $\hV(t)$ at all times during the protocol. This is however an overkill:~to achieve
Eq.~\eqref{eq:PerfectGate} (and hence a perfect evolution), we only need the net effect of leakage
transitions to {\it average away} at the final time. This implies that the final time ``error
propagator'' $\hU_{\mm{err}} = \hU_\uI (t_\uf)$ should essentially act as the unit operator on
computational states, i.e.
\begin{equation}
	\hP_Q \hU_\uI(t_\uf) \hP_Q = \mathbbm{1}_Q,		
	\label{eq:condUI}
\end{equation}

The question then becomes how this cancellation can be achieved by modifying the available control
fields, while still having $\hU_0(t_\uf)$ generate the desired unitary in the computational
subspace. The modification of the control fields changes $\hH(t)$:~this change is described by a
``correction Hamiltonian'' $\hW(t)$, defined via
\begin{equation}
	\hH (t) \to \hH (t) + \hW (t).
	\label{eq:Hctrl}
\end{equation}
We want the combined effect of the correction Hamiltonian $\hW (t)$ and the leakage Hamiltonian
$\hV(t)$ to average away at the final time. The introduction of $\hW(t)$ modifies the error
propagator to be
\begin{equation}
	\begin{aligned}
	\hU_{\mm{err,mod}} &= \hT\exp\left\{-i \int_0^{t_\uf} \ud t_1 \left[\epsilon \hV_\uI (t_1) 
	+ \hW_\uI (t_1)\right]\right\}.
	\end{aligned}
	\label{eq:Uerrmod}
\end{equation}
We thus need to chose $\hW(t)$ so that 
\begin{equation}
	\hP_Q \hU_{\mm{err,mod}} \hP_Q = \mathbbm{1}_Q.
	\label{eq:defHctrl}
\end{equation}

The basic strategy for dealing with leakage is to find (to some desired degree of accuracy) a
method for fulfilling Eq.~\eqref{eq:defHctrl} and thus cancelling on average the effects of leakage.
If successful, the modified protocol will cause the system to depart from the computational subspace
at intermediate times, but return to it at the final time to perform the desired gate [see
Fig.~\ref{fig:idea}~(c)]. 

\section{Magnus approach for constructing corrected control Hamiltonains}
\label{sec:magnussol}

\subsection{Basic approach}
\label{sec:basicapproach}

Finding a suitable $\hW(t)$ to fulfil Eq.~\eqref{eq:defHctrl} remains a daunting problem.
There is however a situation where the solution is straightforward. Suppose that the Hamiltonian
$\hV_\uI (t)$ commuted with itself all times, $[\hV_\uI (t_1), \hV_\uI (t_2)] = 0,\,\forall
t_1,\,t_2$. In this case the time-ordering in Eq.~\eqref{eq:Uerrmod} plays no role, and leakage
errors are completely suppressed whenever $\hW(t)$ satisfies
\begin{equation}
	\int_0^{t_\uf} \ud t_1 \epsilon \sq \hV_\uI  (t_1) 
	= -\int_0^{t_\uf} \ud t_1 \hW_\uI (t_1).
	\label{eq:Hctrlcom}
\end{equation}
The superoperator $\sq$ simply nulls the part of an operator acting only in the leakage
subspace; it is defined via $\sq \hO (t) = \hO (t) - \hP_{N-Q} \hO (t) \hP_{N-Q}$, with $\hP_{N-Q}$
the projector onto the leakage subspace. Eq.~\eqref{eq:Hctrlcom} tells us that the time average of
$\hW(t)$ needs to cancel the relevant parts of the time-averaged leakage operator $\hV_\uI (t)$.

While Eq.~\eqref{eq:Hctrlcom} does not provide a solution to the more general (and standard) case
where  $\hV_\uI (t)$ does not commute with itself at different times, it does provide a means of
attack: we can perturbatively correct the solution for $\hW(t)$ given by Eq.~\eqref{eq:Hctrlcom} to
account for the non-commutativity of $\hV_\uI (t)$, thus yielding a $\hW(t)$ that satisfies
Eq.~\eqref{eq:defHctrl} to a high degree of accuracy.  We will construct $\hW(t)$ as a series,
\begin{equation}
	\hW (t) = \sum_{k=1}^{\infty} \hW_k (t). 
	\label{eq:WSeries}
\end{equation}
where the first term $\hW_1(t)$ satisfies Eq.~\eqref{eq:Hctrlcom} and scales like $\pp$, while the
$k$th term scales like $\pp^k$.  

As we show below, the well-known Magnus expansion \cite{magnus1954,blanes2009} applied to the error
propagator $\hU_\uI (t)$ can be used to construct each term of the series defined in
Eq.~\eqref{eq:WSeries}.  We will show that keeping the first $k$ terms in this series allows one to
cancel all leakage errors to order $\pp^k$. This is obviously beneficial in cases where the relevant
matrix elements of the leakage operator $\hV_\uI (t)$ are small, as a term $\propto \pp^k$ is
necessarily proportional to $k$ powers of $\hV (t)$.  This approach also affords advantages in cases
where the matrix elements of $\hV_\uI (t)$ are not necessarily small, but are rapidly oscillating.

Consider $\hU_{\mm{err}}^{(j)}$, the modified error propagator in the presence of the first $j$
terms in the expansion of the correction Hamiltonian:
\begin{equation}
	\begin{aligned}
		\hV_\uI^{(j)} (t) & = \epsilon \hV_\uI (t) + \sum_{k=1}^j \hW_{k,\uI} (t), \\
		\hU_{\mm{err}}^{(j)} & = \hT\exp\left[-i \int_0^{t_\uf} \ud t_1 \hV_\uI^{(j)} (t_1)\right] .
	\end{aligned}
\label{eq:moderrprop}
\end{equation}
The Magnus expansion expresses these time-ordered exponentials as the simple exponential of an
operator, which is itself defined as an infinite series. Applying it to $\hU_{\mm{err}}^{(j)}$
yields:  
\begin{equation}
	\begin{aligned}
		\hU_{\mm{err}}^{(j)} (t) & = \exp \left[ \hOmega^{(j)}(t) \right],  \\
		\hOmega^{(j)}(t) & = \sum_{k=1}^\infty \hOmega_k^{(j)} (t).
	\end{aligned} 
	\label{eq:moderrpropMagnus}
\end{equation}
The first terms of each series are given by (see, e.g.~\cite{magnus1954,blanes2009})
\begin{equation}
	\begin{aligned}
		\rd_t \hOmega^{(j)}_1 (t) &= -i \hV^{(j)}_\uI (t) ,\\
		\rd_t \hOmega^{(j)}_2 (t) &= \frac{1}{2}\left[\rd_t \hOmega^{(j)}_1 (t), \hOmega^{(j)}_1 (t)\right],\\
		\rd_t \hOmega^{(j)}_3 (t) &= \frac{1}{2}\left[\rd_t \hOmega^{(j)}_1 (t), \hOmega^{(j)}_2 (t)\right] 
		- \frac{1}{6}\left[\hOmega_1^{(j)} (t), \rd_t \hOmega^{(j)}_2 (t)\right],\\
		\ldots
	\end{aligned}
	\label{eq:magnus_series}
\end{equation}
Further, the initial condition $\hU^{(j)}_\uI (0) = \mathbbm{1}$ implies $\hOmega^{(j)}(0) =
\mathbf{0}$.  In what follows, we refer to $\hOmega^{(j)}(t)$ as the ``Magnus operator"
corresponding to the $j$th order corrected error propagator $\hU_{\mm{err}}^{(j)}$.

Leaving the discussion of convergence to Appendix~\ref{sec:magnus_conv}, we first
discuss how to use the above expansions to construct the correction Hamiltonian $\hW(t)$. Our
general strategy is to pick the first $k$ terms in the series for $\hW(t)$ such that we suppress
terms up to order $\pp^k$ in the corresponding error propagator $\hU_{\mm{err}}^{(k)}$. This is
equivalent to requiring that at each order, we chose $\hW_k(t)$ so that the Magnus operators satisfy
\begin{equation}
	\sq \hOmega^{(j)}(t_\uf) \overset{!}{=} \mathcal{O}[\pp^{j+1}] \, \, \, \forall j \leq k.
	\label{eq:OmegakCondition}
\end{equation}

\begin{figure}[t!]
	\includegraphics[width=0.7\columnwidth]{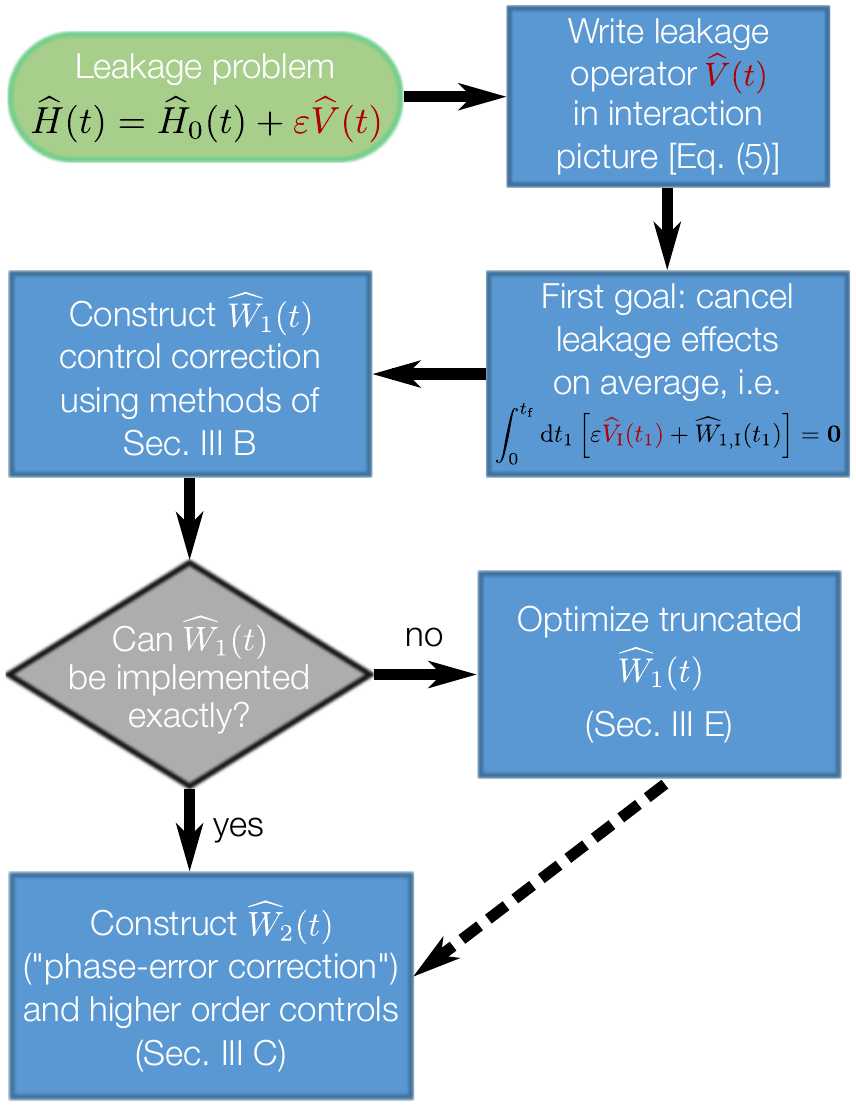}
	\caption{(Color online). Flowchart illustrating the first few steps of the Magnus-based
	algorithm.}
	\label{fig:flowchart}
\end{figure}

\subsection{Constructing the first order correction to the control Hamiltonian}
\label{sec:W1}

We start with the first order term in the correction Hamiltonian $\hW_1(t)$.  To cancel error terms
that are order $\pp$, we require
\begin{equation}
	 \sq \hOmega_1^{(1)} (t_\uf) = \mathbf{0},
	 \label{eq:Hk_choice}
\end{equation}
i.e.~the first order term in the Magnus expansion of the modified error propagator should vanish.
From Eq.~\eqref{eq:magnus_series}, this implies that $\hW_1 (t)$ should satisfy Eq.~\eqref{eq:Hctrlcom},
i.e.
\begin{equation}
	\int_0^{t_\uf} \ud t_1 \pp \sq \hV_\uI (t_1) = -\int_0^{t_\uf} \ud t_1 \hW_{1,\uI} (t_1),	
	\label{eq:Hctrl1gen}
\end{equation}
Note that with this choice $\hW_{1}(t)$ is $\mathcal{O}[\pp]$, and all remaining terms in the Magnus
series for $\sq \hOmega^{(1)}(t_{\uf})$ are order $\mathcal{O}[\pp^2]$ or higher.  These additional
terms also vanish completely in the simple case where $\hV_\uI (t)$ commutes with itself at all
times.

There are of course many possible choices of $\hW_1 (t)$ that will satisfy Eq.~\eqref{eq:Hctrl1gen};
while in principle all such choices are equally as good, in practice not all will be attainable
given the typically limited set of control fields that can be experimentally implemented.  Depending
on the particular problem, one now has two general choices:
\begin{itemize}
	\item  One can attempt to find a $\hW_1 (t)$ that {\it exactly} satisfies
		Eq.~\eqref{eq:Hctrl1gen}.  As we show in Appendix~\ref{sec:fid}, this implies that the
		state-averaged fidelity error scales as $\pp^4$ (as opposed to $\pp^2$ with no
		correction). 
	
	\item  Alternatively, for more complex problems (or where the form of possible control
		Hamiltonians is tightly constrained) one can attempt to satisfy
		Eq.~\eqref{eq:Hctrl1gen} {\it approximately}.  This does not change the power with which
		the fidelity error scales with $\pp$, but can significantly modify the prefactor
		of the leading $\pp^2$ term.  As we show in the Sec.~\ref{sec:applications} with specific
		examples, this can still give significant benefits. This more approximate approach
		is discussed in more detail in Sec.~\ref{sec:imperfectctrls}.  
\end{itemize}
In Fig.~\ref{fig:flowchart}, we depict a flowchart illustrating the first few steps of the
Magnus-based algorithm. 

We outline below two general methods for obtaining first order control corrections $\hW_1(t)$ that
exactly satisfy Eq.~\eqref{eq:Hctrl1gen}.

\subsubsection{Derivative-based control}
\label{sec:W1dt}

One general method for obtaining a $\hW_1(t)$ satisfying Eq.~\eqref{eq:Hctrl1gen} is to use the fact
that in most physically relevant situations, the time dependence in $\hV(t)$ arises from an external
control that is switched on at $t=0$ and off at $t=t_{\uf}$, and hence one often has $\hV(0) =
\hV(t_\uf)= \mathbf{0}$. If this is the case we can always pick $\hW_{1} (t)$ so that the integrand in the
first order Magnus term $\sq \hOmega^{(1)}_1(t)$ is a total derivative, ensuring it vanishes.  

To see this explicitly, we first define the superoperator
\begin{equation}
	\EuScript{L}_0 (t) = \int_0^t \ud t_1 \ell_0(t_1),
	\label{eq:L0}
\end{equation}
where the superoperator $\ell_0(t)$ is defined in Eq.~\eqref{eq:SuperOpL}.  We then choose
\begin{equation}
	\hW_{1} (t) = \pp \ell_0^\dag (t) \EuScript{L}_0 (t) \rd_t \sq \hV(t).
	\label{eq:Hctrl1}
\end{equation}
This yields
\begin{equation}
	\begin{aligned}
		\sq \hOmega_1^{(1)} (t_\uf) &= -i \pp \int_0^{t_\uf} \ud t_1 \ell_0(t_1) \times \\
		&\phantom{={}} 
		\left[\sq \hV (t_1) + \ell_0^\dag (t_1) \EuScript{L}_0 (t_1) \rd_{t_1} \sq \hV (t_1)\right]\\ 
		&= -i \pp \int_0^{t_\uf} \ud t_1 \rd_{t_1} \left[\EuScript{L}_0 (t_1) \sq \hV (t_1)\right]\\
		&= -i \pp \left[\EuScript{L}_0 (t_\uf) \sq \hV(t_\uf) - \EuScript{L}_0 (0)
		\sq \hV(0)\right]
		= \mathbf{0}.
	\end{aligned}
	\label{eq:Omega1}
\end{equation}
Although the form for the correction in Eq.~\eqref{eq:Hctrl1} may seem opaque, we will show in later
sections that it can be explicitly found and implemented in a number of relevant problems. This
solution also provides a good starting point when looking for controls that are constrained by the
physical system into consideration. We refer to Eq.~\eqref{eq:Hctrl1} as the
\textit{derivative-based control}.

The control defined in Eq.~\eqref{eq:Hctrl1} has a striking similarity with DRAG since it involves
the time-derivative of $\hV (t)$. A crucial difference is that $\rd_t \hV (t)$ is multiplied by the
superoperator $\ell_0^\dag (t)\EuScript{L}_0 (t)$, which ensures that $\hW_1 (t)$ is a well-behaved
function of time. This is in contrast to DRAG where the control Hamiltonians may become unphysical
if the unperturbed Hamiltonian is time-dependent.  We give a simple example of this in
Sec.~\ref{sec:STIRAPintro}, for a problem of adiabatic state transfer where the adiabatic gap is
time-dependent.

\subsubsection{Generating function approach}
\label{sec:genfct}

Another extremely general method is to {\it first} pick an operator-valued function $\hR (t)$
that satisfies $\hR (t_f) = \mathbf{0}$. We then can select $\hW_1(t)$ to make the leading-order Magnus
term $\hOmega_1^{(1)} (t) = \hR (t)$. By definition, Eq.~\eqref{eq:Hk_choice} is then satisfied
as required. Using Eq.~\eqref{eq:magnus_series}, the required $\hW_1(t)$ is then found to be:
\begin{equation}
	\begin{aligned}
		\hW_1 (t) &= i \ell_0^\dag (t) \rd_t \sq \hR (t) - \epsilon \sq \hV (t)\\
		&=  i \ell_0^\dag (t) \rd_t \sq \hOmega_1^{(1)} (t) - \epsilon \sq \hV (t).
	\end{aligned}
	\label{eq:Hctrl1Mag1}
\end{equation}
If one can find a class of functions $\hR (t)$ (depending say on a finite number of parameters),
this then yields a family of possible first order control corrections $\hW_1(t)$. Note that since
$\hOmega_1^{(1)} (t)$ is $\mathcal{O}(\epsilon)$, we have that $\hW_1 (t)$ is
$\mathcal{O}(\epsilon)$.

\subsection{Higher-order corrections to the control Hamiltonian}
\label{subsec:HigherOrder}

We now show how to find higher-order control corrections $\hW_{k}(t)$ and in particular the second
order control $\hW_2 (t)$. As for the first order correction, one has considerable freedom in
picking the form of the higher-order controls. This directly follows from
Eq.~\eqref{eq:OmegakCondition} which only requires the Magnus operators to only vanish at $t_\uf$.

Consider first $\hW_{2}(t)$, which suppresses errors at order $\pp^2$. The procedure is completely
analogous to how we chose $\hW_1(t)$: we pick $\hW_{2}(t)$ so that all non-zero terms in $\sq
\hOmega^{(2)}(t_{\uf})$ are at least order $\mathcal{O}[\pp^3]$. As we show explicitly in Appendix
\ref{sec:cancellations}, the needed correction has to fulfill 
\begin{equation}
	-i \int_0^{t_\uf} \ud t_1 \hW_{2,\uI} (t_1) = - \sq \left[\hOmega_2^{(1)}(t_\uf) -
	\hOmega_2^{(1)}(0)\right],
	\label{eq:Hctrl2avg}
\end{equation}
where $\hOmega_2^{(1)}(t)$ is obtained from Eq.~\eqref{eq:magnus_series}. We stress that $\hW_2 (t)
= \ell_0^\dag (t) \hW_{2,\uI} (t)$ is $\mathcal{O}[\pp^2]$ since $\hOmega_2^{(1)} (t)$ is
$\mathcal{O}[\pp^2]$. The superoperator $\ell_0(t)$ is defined in Eq.~\eqref{eq:SuperOpL}.

There is a particular choice of $\hW_2 (t)$ that automatically fulfills Eq.~\eqref{eq:Hctrl2avg}. By
taking the time-derivate (with respect to $t_\uf$) on both sides of Eq.~\eqref{eq:Hctrl2avg}, we
find 
\begin{equation}
	\hW_2 (t) = -i \ell_0^\dag (t) \rd_t \sq \hOmega_2^{(1)} (t).
	\label{eq:Hctrl2}
\end{equation}
This choice for $\hW_2 (t)$ is very convenient as it is fully determined by the form of the leakage
operator and $\hW_{1}(t)$. In particular, Eq.~\eqref{eq:Hctrl2} implies $\sq \hOmega^{(2)} (t) = \sq
\hOmega_1^{(1)} (t) + \mathcal{O}[\pp^3]$. As we have explicitly ensured that $\sq \hOmega_1^{(1)}
(t_{\uf}) = \mathbf{0}$, this immediately implies that Eq.~\eqref{eq:OmegakCondition} is satisfied
as desired.

The recipe for constructing $\hW_k (t)\,\forall k\geq 2$ is the same as the strategy for $k=2$:
we can always choose $\hW_k (t)$ such that $\sq \hOmega^{(k)} (t) = \sq \hOmega_1^{(1)} (t) +
\mathcal{O}[\pp^{k+1}]$, which in turn implies that Eq.~\eqref{eq:OmegakCondition} is satisfied. 

While in most of the examples found in Sec.~\ref{sec:applications} we use Eq.~\eqref{eq:Hctrl2} to
find the second order control Hamiltonian, we illustrate the use of Eq.~\eqref{eq:Hctrl2avg} in
Sec.~\ref{sec:ocs}. 

\subsection{Heuristic interpretation of the controls}
\label{sec:interpretation}

As stated above, keeping only the first two terms in the series for $\hW(t)$ will be sufficient to 
enhance the fidelity; it is thus useful to have a physical picture for these corrections. The
first order Magnus expansion of the uncorrected leakage operator, 
$\hOmega_1^{(0)}$ [see Eq.~\eqref{eq:magnus_series}], represents a time-integrated effective matrix
element for transitions out of the computational subspace $\HHcomp$; the correction $\hW_1(t)$
thus describes an added control that cancels direct leakage errors [up to corrections due to the
non-commutativity of $\hV(t)$].

In contrast, we can understand $\rd_t \hOmega_2^{(1)} (t)$ as the operator that describes the state
of the system when the corrected leakage operator $\hV_\uI^{(1)} (t)$ has acted twice upon it. By
writing $\hV_\uI^{(1)} (t) = \sq V_\uI^{(1)} (t) + \hP_{N-Q} V_\uI^{(1)} (t) \hP_{N-Q}$, we can
identify two sources of error:
\begin{enumerate}[label=\roman*)] 
	\item $[\sq \hV_\uI^{(1)} (t)]^2$ is an operator that acts only within $\HHcomp$,
		but where excursions to the leakage subspace at intermediate times nonetheless yield
		errors. We refer to the latter as ``phase errors''.  
	\item $\{\sq \hV_\uI^{(1)} (t), \hP_{N-Q} V_\uI^{(1)} (t) \hP_{N-Q}\}_+$ is an operator that
		induces leakage out of $\HHcomp$ at order $\epsilon^2$.  We use
		$\{\hO_1,\hO_2\}_+$ to denote the anti-commutator between two operators.  
\end{enumerate}
Thus, the second order correction  $\hW_2 (t)$ describes a control Hamiltonian that corrects both
the phase evolution of states belonging to $\HHcomp$ and cancels leakage errors that scale like
$\mathcal{O}(\epsilon^2)$.

In Figs.~\ref{fig:idea}~(b) and (c), we have depicted the dynamics of the system in the presence of
$\hW (t) = \hW_1 (t)$ and $\hW (t) = \hW_1 (t) + \hW_2 (t)$, respectively.

\subsection{Imperfect realizations of controls}
\label{sec:imperfectctrls}

While the Magnus approach allows one to easily find control corrections for a given problem, in many
cases these corrections will require control fields that are not realizable given experimental
limitations.  In the following, we show that by properly implementing an optimized version of the
truncated control corrections (only using experimentally accessible fields), one can still obtain
significant error suppression.  This statement can be quantified.  If one could perfectly implement
a first order Magnus correction $\hW_1(t)$, the error in the state-averaged fidelity $\bar{F}$
scales like $\epsilon^4$ (see Appendix \ref{sec:fid}).  In the case where $\hW_1(t)$ cannot be
implemented exactly, but instead an optimized, truncated form is used, the fidelity behaves like
$\bar{F} = 1 - a\epsilon^2 + \mathcal{O}(\epsilon^4)$.  While the optimization cannot
eliminate the $\epsilon^2$ term completely, it can make the prefactor $a \ll 1$. 

We consider the general situation where one has derived an ideal first order Magnus correction
$\hW_1(t)$ [i.e.~satisfying Eq.~\eqref{eq:Hctrl1gen}], but where not all of the terms in this
control Hamiltonian can be experimentally achieved.  We thus write the ideal control as
\begin{equation}
	\hW_1 (t) = \hW_1^{\mm{ctrl}} (t) + \hW_1^{\mm{err}} (t),
	\label{eq:partialHctrl1}
\end{equation}
where $\hW_1^{\mm{ctrl}} (t)$ and $\hW_1^{\mm{err}} (t)$ are, respectively, the
experimentally-implementable and unimplementable parts of $\hW_1(t)$. We discuss various
truncation-plus-optimization schemes in what follows.

\subsubsection{Simple truncation}
\label{sec:imperfectctrls_spr}

The simplest strategy would be to simply truncate $\hW_1(t)$ and only implement its
experimentally-attainable part $\hW_1^{\mm{ctrl}} (t)$.  In this case, one no longer completely
cancels the first order term in the Magnus expansion of the error propagator.  The error at $t_\uf$
is fully determined by the Magnus expansion of $\epsilon \sq \hV_\uI (t) + \hW_{1,\uI}^{\mm{ctrl}}
(t)$, which we denote $\hXi^{(1)} (t)$. We have 
\begin{equation}
	\begin{aligned}
		\sq \hXi_1^{(1)} (t_\uf) &= -i \int_0^{t_\uf} \ud t_1 \left[\epsilon \sq \hV_\uI (t_1) +
		\hW_{1,\uI}^{\mm{ctrl}} (t_1)\right]\\
		&= i \int_0^{t_\uf} \ud t_1 \hW_{1,\uI}^{\mm{err}} (t_1),
	\end{aligned}
	\label{eq:errpartialHctrl1}
\end{equation}
which follows from Eqs.~\eqref{eq:Hk_choice} and \eqref{eq:partialHctrl1}. The naive truncation can
be sufficient if one is almost able to implement all of the ideal correction $\hW_1(t)$.  This is
equivalent to requiring 
\begin{equation}
	\int_0^{t_\uf} \ud t_1 \norm{\hW_{1,\uI}^{\mm{err}} (t_1)}_2 \ll  \int_0^{t_\uf} \ud t_1
	\epsilon \norm{\hV_{\uI}(t_1)}_2,
\end{equation}
i.e. the error associated to the modified leakage operator $\hV_\uI (t) + \hW_{1,\uI}^{\mm{ctrl}}
(t)$ is smaller than the error associated to $\hV_\uI (t)$.

\subsubsection{Variational approach}
\label{sec:imperfectctrls_var}

The general goal of our first order correction is to make the time-average of $\epsilon \hV(t) +
\hW_1(t)$ vanish.  While the ideal correction $\hW_1(t)$ accomplishes this task by construction,
simply truncating it causes the integral to no longer be zero, c.f.
Eq.~\eqref{eq:errpartialHctrl1}.  A simple way to try and reduce the magnitude of this integral is
to treat the overall amplitude $\alpha$ of the implementable part of $\hW_1(t)$ as a variational
parameter, i.e. $\hW_1^{\mm{ctrl}} (t) \rightarrow \alpha \hW_1^{\mm{ctrl}} (t)$.  One can then
optimize the value of $\alpha$ to minimize the error integral: 
\begin{equation}
	\begin{aligned}
		\sq \hXi_1^{(1)} (\alpha,t_\uf) &= -i\int_0^{t_\uf} \ud t_1 
		\left[\epsilon \sq \hV_\uI (t_1) + \alpha \hW_{1,\uI}^{\mm{ctrl}} (t_1) \right].
	\end{aligned}
	\label{eq:variationalHctrl1}
\end{equation}
To do this optimization, one needs a suitable scalar measure of the size of this error integral. The
relevant measure here is the state-averaged fidelity $\bar{F}$~\cite{pedersen2007}, which quantifies
how much on average a non-zero $\sq \hXi_1^{(1)} (\alpha,t_\uf)$ causes the evolution under our gate
to deviate from the ideal version. One finds
\begin{equation}
	\bar{F}(\alpha) = \frac{N + \abs{\mm{Tr}\left\{ \exp\left[\sq \hXi_1^{(1)}
	(\alpha, t_\uf)\right] \mathbbm{1}_N\right\}}^2}{N(N+1)}
	\label{eq:avgFopt}
\end{equation}
where $N$ is the dimension of the total Hilbert space. The variational approach then involves
minimizing $\bar{F}(\alpha)$ with respect to $\alpha$. While this variational strategy respects the
ideology of our approach, one could also optimize the state-averaged fidelity for the full error
propagator $\hU_{\mm{err}}^{(1)}$. 

We give an explicit example of this variational approach in Sec.~\ref{sec:gaussian}, in the context
of a quantum state transfer protocol. Also, as discussed more in Sec.~\ref{sec:drag}, our
variational approach sheds insight into the recent work of Chen \textit{et al.}~\cite{chen2016}.
This work discusses how an ad-hoc optimization of prefactors in standard DRAG-style pulse
corrections could help suppress leakage-related errors in superconducting qubit gates; it can also
be viewed as a version of our variational approach to dealing with missing controls.

One could of course do a more sophisticated procedure by optimizing the implementable control fields
at each time.  However, the resulting complexity would defeat the purpose of the Magnus-based
approach, and  would be akin to that of optimal control techniques. In contrast to optimal control,
maximizing $\bar{F}(\alpha)$ is not demanding numerically; nonetheless, it still leads to
significant improvements as compared to the simple truncation strategy.

\subsubsection{Iterative approach}
\label{sec:imperfectctrls_it}

If $\hV (0) = \hV (t_\uf) = \mathbf{0}$, an alternate approach to optimizing the truncated control
Hamiltonian, is to first focus on the form of the unattainable controls $\hW_1^{\mm{err}}(t)$.  As
we care about the \textit{time-integral} of this operator, it is possible in many cases (via
suitable integration by parts) to turn parts of this operator into experimentally attainable
controls. This is especially easy to implement if we have
\begin{equation}
	\rd_t \hW_1^{\mm{err}} (t) = g_{\mm{e}} (t) \hW_1^{\mm{err}} (t) + g_{\mm{c}} (t) \hW_1^{\mm{ctrl}} (t)
\end{equation}
for some functions $g_e(t)$ and $g_c(t)$. If the above condition is fulfilled, then the first order
Magnus expansion of the error propagator (including the ideal first order control correction) can be
written
\begin{equation}
	\begin{aligned}
		&\sq \hOmega_1^{(1)} (t) = 
		-i \int_0^t \ud t_1 \left[\epsilon \sq \hV_\uI (t_1) 
		+ \hW_{1,\uI}^{\mm{ctlr}} (t_1) +
		\hW_{1,\uI}^{\mm{err}} (t_1)\right] \\
		&= -i \int_0^t \ud t_1 \left\{\epsilon \sq \hV_\uI (t_1) + \hW_{1,\uI}^{\mm{ctlr}} (t_1)
		+ \left[\rd_{t_1} \sL_0 (t_1) \right] \hW_1^{\mm{err}} (t_1)\right\} \\
		&= -i \int_0^t \ud t_1 \left[\epsilon \sq \hV_\uI (t_1) + \hW_{1,\uI}^{\mm{ctlr}} (t_1)\right]
		-i\sL_0  (t) \hW_1^{\mm{err}} (t) \\
		&\phantom{={}} 
		+ i \int_0^t \ud t_1 \sL_0 (t_1) \left[ g_{\mm{e}} (t_1) \hW_1^{\mm{err}} (t_1) +
		g_{\mm{c}} (t_1) \hW_1^{\mm{ctrl}} (t_1)\right],
	\end{aligned}
	\label{eq:Xctrl1derivation}
\end{equation}
where the superoperator $\sL_0(t)$ is defined in Eq.~\eqref{eq:L0}.  In the last equality, we have
performed an integration by parts, and used the fact that $\hW_1^{\mm{err}} (0) = \hW_1^{\mm{err}}
(t_\uf) = \mathbf{0}$ which is a consequence of having $\hV (0) = \hV (t_\uf) = \mathbf{0}$. 

Equation~\eqref{eq:Xctrl1derivation} suggests a natural modification of the control Hamiltonian:
$\hW_1^{\mm{ctrl}} (t) \to \hW_1^{\mm{ctrl}} (t) + \hX_1^{\mm{ctrl}} (t)$ with 
\begin{equation} 
	\hX_1^{\mm{ctrl}} (t) = g_{\mm{c}} (t) \ell_0^\dag (t) \sL_0 (t)\hW_1^{\mm{ctrl}} (t).  
	\label{eq:Xctrl1} 
\end{equation}

The error of the new control sequence is determined by 
\begin{equation}
	\sq \hXi_1^{(1)} (t_\uf) = -i \int_0^{t_\uf} \ud t_1 g_{\mm{e}} (t_1) \sL_0 (t_1)
	\hW_1^{\mm{err}} (t_1).
	\label{eq:errorit}
\end{equation}
Generically, for sufficiently slow control protocols, we expect the size of the modified error integral
in Eq.~\eqref{eq:errorit} (as measured e.g. by the $p=2$ operator norm) to be smaller than what we
would have with just a naive truncation, Eq.~\eqref{eq:errpartialHctrl1}.  This is directly due to
the presence of $\sL_0 (t)$, which involves integrating the interaction-picture evolution operator.
It thus roughly scales like the inverse of the energy gap separating computational and leakage
levels. If this gap is large compared to the typical magnitude of $g_\ue (t)$ (which scales like the
inverse timescale of the derivative of the control pulse), our procedure will significantly reduce
the error.

Note that the above procedure represents just the simplest version of a whole hierarchy of control
optimizations. Due to the presence of $\hW_1^{\mm{err}} (t)$ in the last line of
Eq.~\eqref{eq:Xctrl1derivation}, one can repeat the procedure iteratively (i.e.~via additional
integration by parts) to reduce the error even further. Each iterative steps generates a new control
Hamiltonian $\hX_k^{\mm{ctrl}} (t)$ ($k\geq 2$). In Sec.~\ref{sec:drag}, we give an explicit example
of implementing this procedure, in the context of fighting leakage errors in a superconducting qubit
gate. We show that the approach can yield advantages over standard DRAG corrections.

\subsubsection{Second- and higher-order control Hamiltonians}
\label{sec:W2approx}

The previous discussion provided several approaches for dealing with the common case where the ideal
first order control-correction Hamiltonian cannot be realized using the available experimental
controls.  Our general strategy relies in ``truncating''  $\hW_1 (t)$ and subsequently in optimizing
the truncated control Hamiltonian $\hW_1^{\mm{ctrl}} (t)$. As mentioned earlier, this procedure will
lead to a state-averaged fidelity whose leading order correction is proportional to
$\epsilon^2$.

Even though one has not completely cancelled the $\epsilon^2$ term in the error, there is still in
general utility to finding higher-order control correction Hamiltonians as one would do in the ideal
case (c.f.~Sec.~\ref{subsec:HigherOrder}).  This is perhaps best understood physically, using the
heuristic interpretation of the control corrections introduced in Sec.~\ref{sec:interpretation}.
Recall that in that section, we discussed how the principal role of the first order control
correction $\hW_1(t)$ is to cancel pure leakage errors (where the final state lies outside the
computational subspace $\HHcomp$), while the second order correction $\hW_2(t)$ corrects phase
errors (i.e.~helps make sure that the portion of the final state in $\HHcomp$ has the correct form).
In general, cancelling phase errors is useful even if one has not completely cancelled pure leakage
errors by implementing the ideal first order correction $\hW_1(t)$.

A further reason for attempting to implement the higher-order controls is that they are very easily
obtained from the first order control (exactly like the ideal case). In particular, the second order
control is given by Eq.~\eqref{eq:Hctrl2} with $\sq \hOmega_2^{(1)}(t)$ replaced by $\sq
\hXi_2^{(1)} (t)$. We show the utility of this approach in the examples of suppressing non-adiabatic
errors in STIRAP with a time-dependent gap (Sec.~\ref{sec:gaussian}) and correcting superconducting
qubit gates (Sec.~\ref{sec:drag}). 

\section{Applications}
\label{sec:applications}

In this section, we apply our general method to a few concrete problems.
These highlight the flexibility and utility of the method, and also allow one to understand better
similarities and differences from existing methods such as STA and DRAG.

\subsection{Suppressing non-adiabatic errors in STIRAP quantum state transfer}
\label{sec:STIRAPintro}

We start with the problem of adiabatic quantum state transfer.  Consider a $\Lambda$-system with two
degenerate ground states $\ket{1}$ and $\ket{3}$, and an excited state $\ket{2}$. We assume (as is
standard) that states $\ket{1}$ and $\ket{3}$ can be controllably coupled to $\ket{2}$, but cannot
be coupled directly to one another.  The generic control Hamiltonian thus has the form:
\begin{equation}
	\hH (t) = G_\up (t) \ketbra{1}{2}+ G_\us (t) \ketbra{2}{3} + \mm{H.c.},
	\label{eq:lambdasysH}
\end{equation}
where $G_\up (t)$ and $G_\us (t)$ denote the pump and Stokes pulse amplitudes, respectively. 

Stimulated Raman adiabatic passage (STIRAP)~\cite{bergmann1998,vitanov2001,vitanov2016} is a
technique that allows one to perform adiabatic state transfer from $\ket{1}$ to $\ket{3}$ without
ever occupying $\ket{2}$. Since the technique is based on adiabatic passage, it is intrinsically
slow.  There is thus interest in finding methods to speed up this approach.

Here, we use our Magnus-based algorithm to find control corrections that speed up STIRAP while
keeping a high-fidelity (and while respecting the constraint of no direct coupling between $\ket{1}$
and $\ket{3}$).  Apart from its physical relevance to a variety of systems, the STIRAP problem is
interesting because so-called ``shortcuts to adiabaticity'' (STA)
methods~\cite{demirplak2008,ibanez2012,torrontegui2013} can be used here to find control corrections
allowing a perfect transfer fidelity \cite{baksic2016}.  We will use these to benchmark the
performance of the Magnus method to gain further insight into the underlying physics.  We stress
that for more complex problems (like that in Sec.~\ref{sec:mclz}), STA methods are essentially
impossible to implement, whereas the Magnus approach remains workable and useful.  The STIRAP
problem is also interesting as it highlights how the Magnus method is able to surpass limitations of
the standard DRAG technique for fighting leakage~\cite{gambetta2011}.  

\subsubsection{Correcting constant-gap pulses}
\label{sec:STIRAPconstgap}

We start by considering optimal pulses which keep the instantaneous gap between adiabatic
eigenstates constant.  Using the pulse shape studied extensively in~\cite{vasilev2009}, the control
pulses are written
\begin{equation}
	\begin{aligned}
	G_\up (t) & = G_0 \sin\left[\theta(t)\right], \\
	G_\us (t) &= G_0 \cos\left[\theta(t)\right], \\
	\theta(t) & = \frac{\pi}{2} \frac{1}{1 + \exp[-\nu t]}
	\end{aligned}
	\label{eq:pumpandstokes}
\end{equation}
Here, $G_0$ is the maximal strength of the coupling and $1/\nu$ determines the characteristic
timescale of the protocol. 

\begin{figure}[t!]
	\includegraphics[width=\columnwidth]{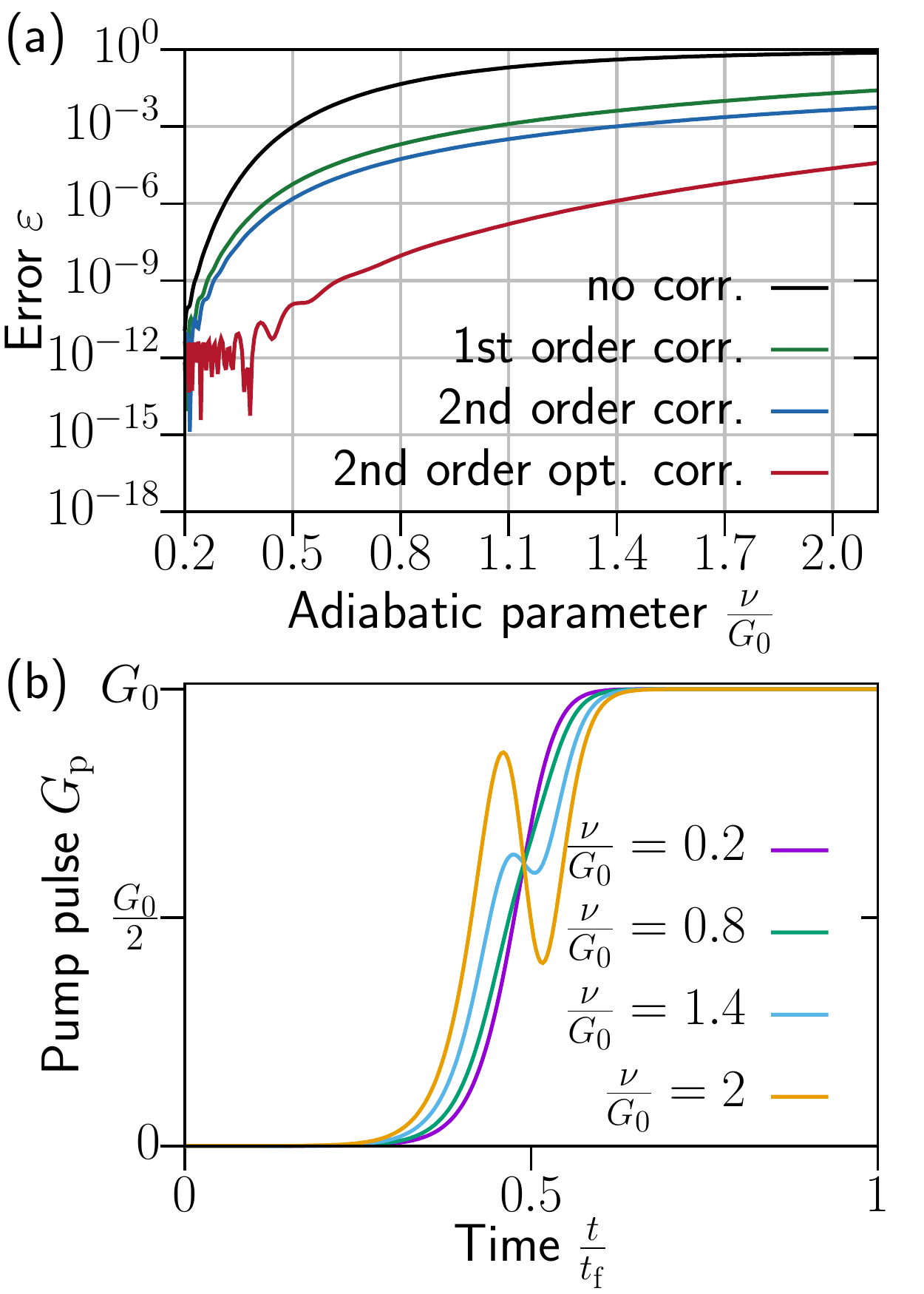}
	\caption{(Color online). (a) Fidelity error for STIRAP-style adiabatic state transfer
		in a $\Lambda$-system (c.f.~Eq.~(\ref{eq:lambdasysH})), as a
		function of the adiabatic parameter $\nu/G_0$.  Different curves correspond to
		different choices of control corrections: ``uncorrected'' control sequence defined
		in Eq.~\eqref{eq:pumpandstokes} (black), first order Magnus corrected controls
		(green), second order Magnus corrected controls (blue), and the optimal second
		order Magnus corrected controls (red).  (b) Modified pump pulse $G_\up (t)$
		predicted by the second order Magnus algorithm as function of $t/t_\uf$ for
		different values of $\nu/G_0$. As $\nu/G_0 \to 0$, the modified sequence converges
		to the original one. The Stokes pulse $G_\us (t)$ is the time-reversed of the pump pulse.}
	\label{fig:fidlambda} 
\end{figure}

We next move to the adiabatic frame.  The Hamiltonian takes the form
\begin{equation}
	\begin{aligned}
	\hH_{\mm{ad}} (t) &=\hS^\dag (t) \hH  (t)  \hS (t) 
	- i \hS^\dag (t) \rd_t \hS (t)  \\
	&=G_0 \left(\ketbra{\mm{b}_+}{\mm{b}_+}-\ketbra{\mm{b}_-}{\mm{b}_-}\right) \\
	&\phantom{=}+\frac{\dot{\theta}(t)}{\sqrt{2}}\left[i\left(\ketbra{\mm{d}}{\mm{b}_+} +
	\ketbra{\mm{d}}{\mm{b}_-}\right) + \mm{H.c.}\right]\\
	&= \hH_0 + \hV (t).
	\end{aligned}
	\label{eq:Hadlambda}
\end{equation}
Here $\hS^\dag(t)$ is the unitary change of basis operator that diagonalizes $\hH(t)$, and maps
states in the lab frame onto states in the adiabatic frame.  We have denoted the adiabatic
eigenstates of the system by $\ket{\mm{b}_+}$, $\ket{\mm{b}_-}$ (known as bright-states), and
$\ket{\mm{d}}$ (dark-state). The corresponding instantaneous eigenenergies are time-independent and
given by $E_{\pm} = \pm G_0$ and $E_{\mm{d}} = 0$. 

An ideal STIRAP-style state transfer would involve having the system stay in the $\ket{\mm{d}}$
state as it adiabatically evolves from being $\ket{1}$ to $\ket{3}$ (accomplished by having
$\theta(t)$ evolve continuously from $0$ to $\pi/2$).  Leakage errors here are due to non-adiabatic
transitions between the adiabatic eigenstates.  They are caused by the term $\hV (t) \propto
\dot{\theta}$ in $\hH_{\mm{ad}}$.  Generically, the size of $\hV(t)$ (and hence leakage errors)
increases as the protocol is made faster.  

The goal as always is to find modifications of the original pulse sequence so as to suppress leakage
(i.e.~non-adiabatic) errors.  To do this using our Magnus approach, we first write $\hV(t)$ in the
interaction picture defined by Eq.~\eqref{eq:SuperOpL} [i.e.~generated by the time-independent
$\hH_0$ in Eq.~\eqref{eq:Hadlambda}]:
\begin{equation}
		\hV_\uI (t) =\frac{\dot{\theta}(t)}{\sqrt{2}}\left[i\left(\ue^{-i G_0 t}
		\ketbra{\mm{d}}{\mm{b}_+}+ \ue^{i
		G_0 t}\ketbra{\mm{d}}{\mm{b}_-} \right) + \mm{H.c.} \right].
	\label{eq:VintMagnus}
\end{equation}
Note that for this problem we have $\sq \hV_\uI (t) = \hV_\uI (t)$ (i.e.~it does not have matrix
elements directly coupling leakage levels to one another). 

The first step of the Magnus approach for deriving control corrections is to find a first order
control correction $\hW_1(t)$ satisfying the cancellation condition of
Eq.~\eqref{eq:Hctrl1gen}.  Since we have $\dot{\theta}(0) = \dot{\theta}(t_\uf) = 0$, we can use the
\textit{derivative-based control} choice for $\hW_1 (t)$ in Eq.~\eqref{eq:Hctrl1}.  Writing the
correction in the adiabatic frame [i.e.~the same frame as Eq.~\eqref{eq:Hadlambda}], we find:
\begin{equation}
	\hW_{1,\mm{ad}} (t) = \frac{1}{\sqrt{2}}\frac{\ddot{\theta}(t)}{G_0}\left[\ketbra{\mm{d}}{\mm{b}_-} -
	\ketbra{\mm{d}}{\mm{b}_+} + \mm{H.c.}\right].
	\label{eq:H1ad}
\end{equation}
As always, we stress that the choice of $\hW_1(t)$ is not unique.  The above choice has the merit of
both being simple to find, and (as one can easily verify) of not requiring any direct coupling
between levels $\ket{1}$ and $\ket{3}$ to implement.

The second order control is fully determined by $\hV_\uI (t)$ and $\hW_{1} (t)$; it can be found
using Eq.~\eqref{eq:Hctrl2}. We find 
\begin{equation}
	\hW_{2,\mm{ad}} (t) = \frac{1}{2}\frac{\dot{\theta}^2 (t)}{G_0}\left(-\ketbra{\mm{b}_+}{\mm{b}_+}
	+ \ketbra{\mm{b}_-}{\mm{b}_-}\right).
	\label{eq:H2ad}
\end{equation}
where we have written the control in the adiabatic frame.  Again, this control Hamiltonian does not
require any direct $\ket{1}$ and $\ket{3}$ coupling.  Hence, implementing both the first and second
order Magnus-derived corrections corresponds to a simple modification of the control pulses $G_{\rm
s}(t)$ and $G_{\rm p}(t)$.  The form of the modified $G_{\mm{p}}(t)$ control pulse is shown in
Fig.~\ref{fig:fidlambda}~(b) for various choices of the speed parameter $\nu/G_0$; $G_{\mm{s}}(t)$
is just given by $G_{\mm{p}}(-t)$.  As $\nu/G_0 \rightarrow 0$, one is in the adiabatic limit, and
the correction to the control pulses vanishes as expected.

To assess the efficacy of the different control sequences, we consider the state infidelity for a
STIRAP style population transfer from $\ket{1}$ to $\ket{3}$.  This is defined as
\begin{equation}
	\varepsilon = 1-\abs{\bra{3} \hU^{(j)}(t_\uf) \ket{1}}^2.
	\label{eq:statefid}
\end{equation}
The behaviour of this error for the uncorrected STIRAP protocol versus the Magnus-based corrected
protocols are shown in Fig.~\ref{fig:fidlambda}~(a).  One sees that if the goal is an error less than
$10^{-3}$\footnote{This threshold is motivated by the requirement for fault-tolerant universal quantum
computing~\cite{raussendorf2007} since STIRAP can be used to implement a $\mm{NOT}$ gate},
the second order Magnus-based correction allows for a $\sim 2.6$-fold increase in the maximum
protocol speed (i.e.~speed parameter $\nu$). Note that to simulate a realistic experiment where
pulses all have a finite duration, we have integrated the Schrödinger equation from $t_\ui = -\ln(-1
+ \pi/2\arcsin[\delta])/\nu$ to $t_\uf = -\ln(-1 + \pi/2\arccos[\delta])/\nu$. This choice ensures
that the control fields are small at both the initial and final protocol times:
$\sin[\theta (t_\ui)] = \cos[\theta (t_\uf)] = \delta$.  We take $\delta = 10^{-6}$.

\subsubsection{Link to shortcuts to adiabaticity}
\label{sec:linkSATD}

As previously discussed, for the STIRAP problem, one can use the STA approach to derive an exact control
correction that always yields perfect fidelity, the  so-called ``super-adiabatic transitionless
driving'' (SATD)~\cite{baksic2016}. These ideas were recently tested
experimentally~\cite{du2016,zhou2016}.  The added control (written in the adiabatic basis) takes the
form:
\begin{equation}
	\hW_{\mm{SATD},\mm{ad}} (t) = \frac{\ddot{\theta} (t)}{\sqrt{2} G_0} \frac{1}{1 +
	\frac{\dot{\theta}^2 (t)}{G_0^2}} 
	\left( \ketbra{\ud}{\ub_-} -\ketbra{\ud}{\ub_+} + \mm{H.c.}\right).
	\label{eq:satdconstgap}
\end{equation}

To compare this correction against those found from our Magnus approach, we can formally expand it
in powers of $\dot{\theta}(t)/G_0$. The zeroth-order term in this expansion is identical to the first
order correction $\hW_1(t)$ found using our Magnus approach, c.f.~Eq.~\eqref{eq:H1ad}.  This
suggests that the Magnus approach provides a perturbative route for obtaining the benefits of a STA
protocol (at least to leading order).  

If one continues to expand the SATD correction in powers of $\dot{\theta}$, one easily sees that the
correspondence to the Magnus approach does not extend past leading order (see
Appendix~\ref{sec:compsatd} for more details).  Recall that the basic physical picture underlying
both approaches is the same, and corresponds to Fig.~\ref{fig:idea}: to fight leakage, one
constructs a ``dressed'' dark state, and implements control corrections to ensure the system
evolution follows this dressed state.  We see that the dressed states for the two approaches are in
general not the same (though they are similar to leading order).

For the STIRAP problem, the existence of the perfect SATD correction makes the Magnus approach
unnecessary.  The same is not true for more complex systems, where STA approaches are impossible to
implement, whereas the Magnus approach remains tractable.  An explicit example of such a problem is
given in Sec.~\ref{sec:mclz}.

\subsubsection{Optimal second order control correction}
\label{sec:ocs}

Given the freedom one has to pick $\hW_2 (t)$ [see Eq.~\eqref{eq:Hctrl2avg}], we can look for an
``optimal'' control, $\hW_{2,\mm{opt}} (t)$, that simultaneously cancels all leakage errors at order
$\epsilon^2$ and $\epsilon^3$. To find such a control, one needs to (numerically) solve a set of
coupled first order differential equations, which defeats the purpose of our Magnus-based method
(see Appendix~\ref{sec:optctrlSTIRAP}).

However, since for STIRAP \textit{pure} leakage errors are more detrimental than pure phase errors,
one can try to find an approximate optimal control that suppresses pure leakage errors at order
$\varepsilon^3$ (i.e third order processes that take the system out of the dark state subspace) at
the expense of not fully cancelling phase errors at order $\varepsilon^2$ (hence giving an
approximate optimal control Hamiltonian). We find (see Appendix~\ref{sec:optctrlSTIRAP})
\begin{equation}
	\hW_{2,\mm{opt},\mm{ad}} (t) \simeq \frac{1}{3}\frac{\dot{\theta}^2(t)}{\Omega_0}
	\left(-\ketbra{\ub_+}{\ub_+} + \ketbra{\ub_-}{\ub_-}\right).
	\label{eq:Hoptad}
\end{equation}
Remarkably, implementing Eq.~\eqref{eq:Hoptad} only requires changing the numerical prefactor of
$\hW_2 (t)$ [see Eq.~\eqref{eq:H2ad}].

In Fig.~\ref{fig:fidlambda}~(a), we have plotted the error obtained with the optimal second order
Magnus-based control (red trace). As it can be seen, the error is always smaller than $10^{-3}$. If
we define the range of validity of a ``corrected'' control sequence as the parameter regime for
which the maximal amplitude of the corrected pulse does not exceed the maximal amplitude of the
original pulse, then the validity of the optimal second order correction is very close to that of
SATD.  We numerically find that the optimal control obeys this criterion for $\nu/\Omega_0 \lesssim
2.21$, while SATD is valid for $\nu/\Omega_0 \lesssim 2.63$.  Within our definition of high-fidelity
adiabatic passage ($\varepsilon \leq 10^{-3}$), the approximate optimal control found at second
order is nearly as good as the exact control.

\subsubsection{Corrections for a time-dependent gap and connections to DRAG}
\label{sec:gaussian}

As previously discussed, the DRAG technique~\cite{motzoi2009,gambetta2011} is also a perturbative technique for
constructing corrections to control pulses to suppress the deleterious effects of leakage
transitions.  The STIRAP problem considered in this section is an excellent testbed for seeing how
it compares to our Magnus approach.  For the uncorrected pulse sequence of the previous subsection
[c.f.~Eq.~\eqref{eq:pumpandstokes}], one finds that the DRAG approach yields identical control
corrections as the Magnus approach, i.e.~$\hW_1(t)$ and $\hW_2(t)$ in Eqs.~\eqref{eq:H1ad} and
\eqref{eq:H2ad}. This is however a special case where the adiabatic gap in the original pulses is
completely time-dependent.  As we now show, for more general pulses, the DRAG technique can fail
completely, while the Magnus approach remains effective.

Consider the most standard pulse shapes used for STIRAP, where both $G_{\mm{p}}(t)$ and $G_{\mm{s}}(t)$ are
Gaussians~\cite{bergmann1998}.  We have explicitly:
\begin{equation}
	\begin{aligned}
	G_\up (t) &= G_0 \exp\left[ -\nu^2 (t - t_0 - \tau)^2\right],\\
	G_\us (t) &= G_0 \exp\left[ -\nu^2 (t-t_0)^2\right],
	\end{aligned}
	\label{eq:gaussian}
\end{equation}
with $\tau$ the delay between the sequences and $t_0 = \sqrt{-\ln(\delta)}/\nu$ is defined such that
$G_\us (0) = \delta$. In the adiabatic frame, the Hamiltonian reads
\begin{equation}
	\begin{aligned}
		\hH_{\mm{ad}} (t) &= G (t) \left(\ketbra{\mm{b}_+}{\mm{b}_+}-\ketbra{\mm{b}_-}{\mm{b}_-}\right) \\
		&\phantom{=}+\frac{\dot{\theta}(t)}{\sqrt{2}}\left[i\left(\ketbra{\mm{d}}{\mm{b}_+} +
		\ketbra{\mm{d}}{\mm{b}_-}\right) + \mm{H.c.}\right]\\
		&= \hH_0 (t) + \hV (t),
	\end{aligned}
	\label{eq:HlambdaadG}
\end{equation}
where $G (t) =  \sqrt{G_\up^2 (t) + G_\us^2 (t)}$ and $\theta (t) = \arctan[G_\up (t) / G_\us
(t)]$. 

\begin{figure}[t!]
	\includegraphics[width=\columnwidth]{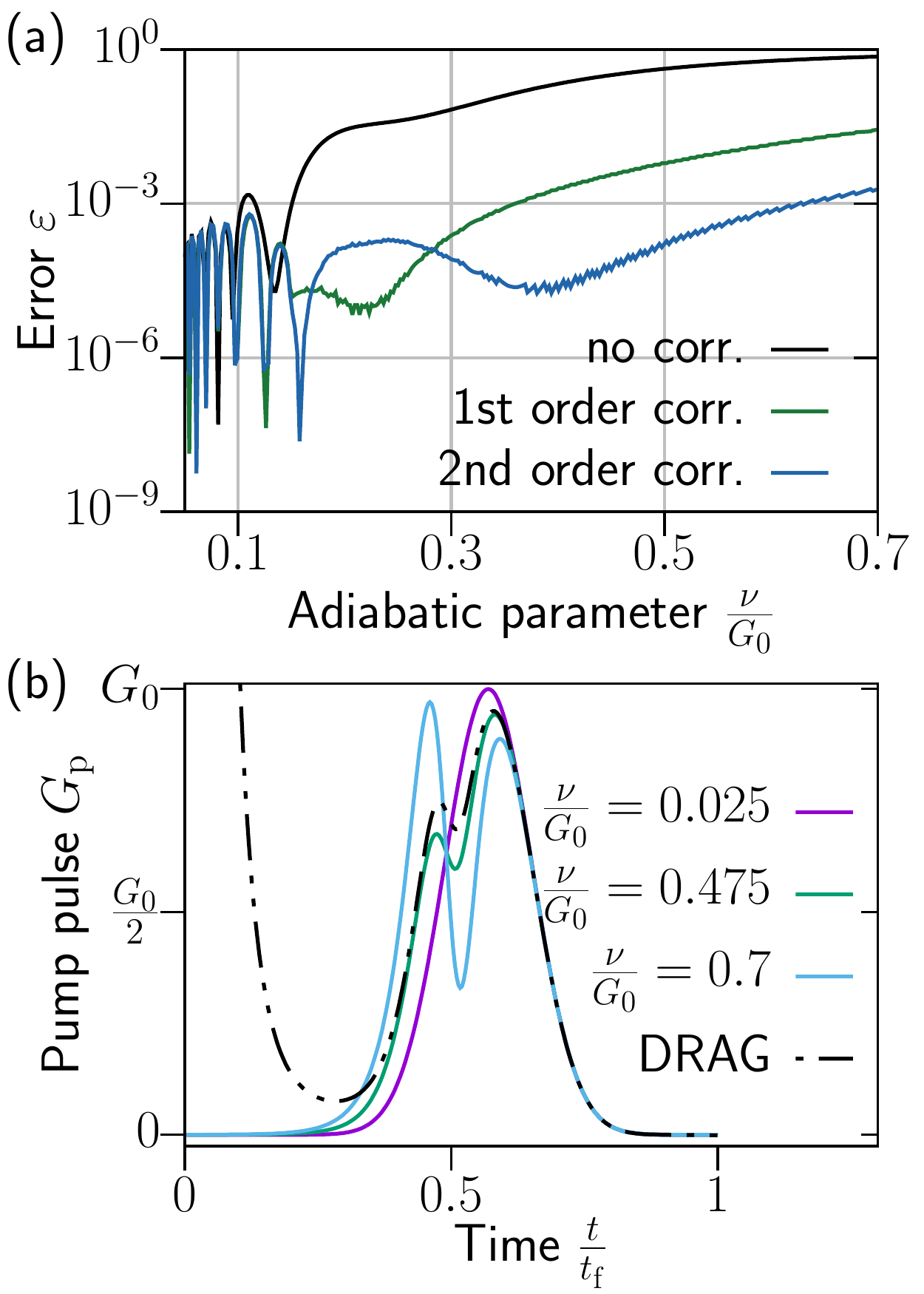}
	\caption{(Color online). (Color online). (a) Fidelity error for STIRAP-style state transfer,
		as a function adiabatic parameter $\nu/G_0$, for uncorrected pulses having a simple
		Gaussian form.  Curves correspond to different choice of correction: ``uncorrected''
		control sequence defined in Eq.~\eqref{eq:gaussian} (black), first order Magnus
		corrected controls (green), and second order Magnus corrected controls (blue).  (b)
		Modified pump pulse $G_\up (t)$ predicted by the second order Magnus algorithm as
		function of $t/t_\uf$ for different values of $\nu/G_0$. As $\nu/G_0 \to 0$, the
		modified sequence converges to the original one. The Stokes pulse is the
		time-reversed of the pump pulse. The black dashed line is the control predicted by
		DRAG for $\nu/G_0 = 0.475$. As one can observe, this control diverges when $t\to 0$.}
	\label{fig:fidgaussian} 
\end{figure}

One can straightforwardly apply standard DRAG to this problem to help cancel non-adiabatic errors.
It predicts a first order correction similar to Eq.~\eqref{eq:H1ad} with $G_0$ replaced by $G (t)$.
The amplitude of the correction is thus $\propto \ddot{\theta} (t) /G(t)$.  Unfortunately, the gap
$G(t) \to 0$ at both late and early times, and does so faster than $\ddot{\theta}(t)$.  As a result
and as shown in Fig.~\ref{fig:fidgaussian}~(b), the DRAG correction diverges, and is thus not
physical.  In contrast, using the Magnus approach yields physical (non-divergent) corrections.  The
leakage Hamiltonian in the interaction picture is:
\begin{equation}
	\hV_\uI (t) =\frac{\dot{\theta}(t)}{\sqrt{2}}
	\left[i\left(e^{-i \Delta (t)}\ketbra{\mm{d}}{\mm{b}_+} 
	+ e^{i \Delta (t)} \ketbra{\mm{d}}{\mm{b}_-} \right) + \mm{H.c.} \right],
	\label{eq:VIG}
\end{equation}
where $\Delta (t) = \int_0^t \ud t_1 G (t_1)$. 

As usual, the first step of the Magnus approach is to construct a first order control correction
$\hW_1(t)$ satisfying Eq.~\eqref{eq:Hctrl1gen}.  While one could again use the
\textit{derivative-based control} technique to find a well-behaved $\hW_1 (t)$, for illustrative purposes,
we will use some of the other techniques discussed in Secs.~\ref{sec:genfct} and
\ref{sec:imperfectctrls}.  In particular, we will use the \textit{generating function approach}
combined with the \textit{variational approach}.

The generating function approach to finding $\hW_1(t)$ requires one to first construct a well-chosen
time-dependent generating function that vanishes at $t=t_f$ [i.e.~satisfies
Eq.~\eqref{eq:Hk_choice}]. Motivated by the form of $\hOmega_1^{(1)} (t)$ that was found in the last
subsection for the case where $G(t)$ was time-independent, we consider the function
\begin{equation}
	\begin{aligned}
		&\hR(\alpha, t) = \hOmega_1^{(1)} (\alpha, t) = \\
		&-\frac{i}{\sqrt{2}} \alpha\left[-e^{-i \Delta (t)} \frac{\dot{\theta}(t)}{G_0}
		\ketbra{\mm{d}}{\mm{b}_+} +  e^{i \Delta (t)} \frac{\dot{\theta}(t)}{G_0}
		\ketbra{\mm{d}}{\mm{b}_-} + \mm{H.c.}\right],
	\end{aligned}
	\label{eq:H1GaussianEq}
\end{equation}
Here, $\alpha$ is a parameter that will be set in what follows.

Given this generating function, the corresponding control-correction $\hW_1(t)$ is found by using
Eq.~\eqref{eq:Hctrl1Mag1}.  Unfortunately, when written in the lab-frame, one sees that it has terms
requiring a direct coupling between levels $\ket{1}$ and $\ket{3}$. As described in
Sec.~\ref{sec:imperfectctrls}, we can still proceed even though $\hW_{1}(t)$ cannot be fully
implemented given our limited  experimental controls. We first decompose it into its attainable
[$\hW_{1,\mm{ad}}^{\mm{ctrl}} (t)$] and unattainable [$\hW_{1,\mm{ad}}^{\mm{err}} (t)$] parts,
$\hW_{1,{\mm{ad}}}(t) = \hW_{1,\mm{ad}}^{\mm{ctrl}} (t)+  \hW_{1,\mm{ad}}^{\mm{err}} (t) $, see
Eq.~\eqref{eq:partialHctrl1}. We find
\begin{equation}
	\hW_{1,\mm{ad}}^{\mm{ctrl}} (t) = \alpha \frac{\ddot{\theta}(t)} {\sqrt{2} G_0}
	\left[ \ketbra{\mm{d}}{\mm{b}_-} - \ketbra{\mm{d}}{\mm{b}_+}  + \mm{H.c.}\right],
	\label{eq:H1Gaussianctrl}
\end{equation}
and 
\begin{equation}
	\hW_{1,\mm{ad}}^{\mm{err}} (t) = \frac{1}{\sqrt{2}} \gamma (t) 
	\left[ i \ketbra{\mm{d}}{\mm{b}_-} + i \ketbra{\mm{d}}{\mm{b}_+}  + \mm{H.c.}\right].
	\label{eq:H1Gaussianerr}
\end{equation}
Here, $\gamma (t) = (\alpha G(t)/G_0 -1)\dot{\theta} (t)$. 

To proceed, we will use the variational approach discussed in Sec.~\ref{sec:imperfectctrls_var}.  We
basically assume that only the correction $ \hW_{1,\mm{ad}}^{\mm{ctrl}} (t)$ is implemented, and
pick $\alpha$ to minimize the error that results from dropping $\hW_{1,\mm{ad}}^{\mm{err}} (t)$.
This involves numerically minimizing the average fidelity $\bar{F}[\alpha]$ defined in
Eq.~\eqref{eq:avgFopt}. In our case, this function takes the simple form:
\begin{equation}
	\bar{F}[\alpha] = \frac{1}{4} + \frac{1}{12}\left[1+ 2\cos\left( \abs{\int_0^{t_\uf} \ud t_1
	\exp[i \Delta (t_1)] \gamma (t_1)} \right)\right]^2.
	\label{eq:avgFoptG}
\end{equation}

Having found the first order control, the second order control Hamiltonian is obtained via
Eq.~\eqref{eq:Hctrl2}. We find
\begin{equation}
	\hW_{\mm{ad},2} (t) = \left(\alpha\frac{\dot{\theta}^2 (t)}{2 G_0} + \beta
	(\alpha, t)\right)\left(\ketbra{\mm{b}_-}{\mm{b}_-} - \ketbra{\mm{b}_+}{\mm{b}_+}\right),
	\label{eq:H2Gaussian}
\end{equation}
with
\begin{equation} 
	\begin{aligned}
	\beta (\alpha, t) &= \mm{Im}\left[\frac{e^{-i \Delta (t)} \dot{\theta}(t)}{2} \int_0^t \ud t_1
	e^{i \Delta (t_1)} \gamma (t_1)\right]\\
	&\phantom{=} + \mm{Re}\left[\alpha \frac{e^{-i \Delta (t)} \ddot{\theta}(t)}{2 G_0} \int_0^t \ud t_1
	e^{i \Delta (t_1)} \gamma (t_1)\right].
	\end{aligned}
	\label{eq:H2Gaussainerrcorr}
\end{equation}

In Fig.~\ref{fig:fidgaussian}~(a), we have plotted the infidelity $\varepsilon$ [see
Eq.~\eqref{eq:statefid}] as a function of the adiabatic parameter $\nu/\Omega_0$. We consider
various controls: the original, uncorrected Gaussian pulses , the first order Magnus correction ,
and second order Magnus correction. For each value of $\nu/\Omega_0$, we have chosen the $\alpha$
that minimizes Eq.~\eqref{eq:avgFoptG}. The exact time evolution operator $\hU^{(j)} (t)$ is found
by integrating the Schrödinger equation from $t = 0$ to $t_\uf$ with $G_\up (0) = G_\us (t_\uf) =
\delta$. We have chosen $\delta = 10^{-6}$. If we consider, as previously, that a high-fidelity
adiabatic passage is achieved for $\log_{10} [\varepsilon^{(j)}] \leq -3$, then the modified control
sequence defined by $\hW (t) = \hW_1 (t) + \hW_2 (t)$ allows for a five-fold speedup compared to the
original Gaussian-based STIRAP sequence. In Figs.~\ref{fig:fidgaussian}~(b) and (c), we plot the
modified pump and Stokes pulses, respectively.  As $\nu/G_0 \to 0$, the modified control sequence
converges to the original sequence. 

The curves in Fig.~\ref{fig:fidgaussian}~(a) illustrate a general important point:  in the extreme
adiabatic limit, $\nu/G_0 \ll 1$, the fidelity error does not tend to zero (as one would naturally
expect), but instead shows an oscillating behaviour.  The errors in this limit are not due to
non-adiabatic transitions, but instead due to the finite protocol duration (see
Appendix~\ref{sec:oscFG}).  By averaging the fidelity error of the uncorrected protocol between
$\nu/G_0 = 0.025$ and $\nu/G_0 = 0.1$, we see that these finite-time errors set an approximate lower
limit on the error of $\sim 10^{-4}$. As our correction protocol is only designed to suppressed
non-adiabatic errors, it will not be able to suppress the total error appreciably below this level.
This also explains why the second-order corrected protocol appears to have a higher error than the
first-order corrected protocol in the range $0.17 \lesssim \nu/G_0 \lesssim 0.28$.  In this regime,
the errors remaining after the application of the first order control are dominated by finite-time
effects, and hence applying a higher-order correction (designed to fight non-adiabatic errors) will
not improve matters.

\subsection{Leakage in a superconducting qubit} 
\label{sec:drag}

The DRAG technique for correcting pulses was originally formulated to suppress leakage during
manipulations of superconducting qubits~\cite{motzoi2009}.  The qubit here is formed by the lowest
two levels of an anharmonic oscillator (eigenstates $\ket{n}$).  Applying our Magnus-based method to
this problem yields a different approach for constructing control corrections.  One immediately
encounters the situation discussed in Sec.~\ref{sec:imperfectctrls}, where the theoretically-ideal
control corrections must be truncated and optimized to reflect the limited experimental control
fields available.  This provides a direct and very physical way to understand the recent
experimental work of Chen \textit{et al.} in Ref.~\cite{chen2016}.  This work demonstrated that an ad-hoc
optimization of DRAG results in significant fidelity improvements.  Our approach shows how the need
for such an optimization arises naturally.  

We consider a situation where a drive is applied to the qubit with a centre frequency
$\omega_{\mm{dr}}$ and complex envelope $2 \kappa(t)$.  In the frame rotating at the drive frequency
(and within the rotating wave approximation), the Hamiltonian describing the system is
\begin{equation} 
	\begin{aligned} 
		\hH (t) &= \frac{\delta}{2} \left(-\ketbra{0}{0} +
		\ketbra{1}{1}\right) + \left(\frac{3}{2}\delta + \Delta\right) \ketbra{2}{2} \\
		&\phantom{=} 
		+ \left(\kappa (t) \ketbra{0}{1} + \lambda \kappa (t) \ketbra{1}{2} + \mm{H.c.}\right).  
	\end{aligned}
	\label{eq:DRAGHamiltonian} 
\end{equation}
$\delta$ is the detuning between the drive frequency and qubit transition frequency, $\Delta$ is the
qubit anharmonicity, and $\lambda$ is a parameter that characterizes the strength of the transition
$\ket{1} \leftrightarrow \ket{2}$ relative to the transition $\ket{0} \leftrightarrow \ket{1}$.  We
take the (uncorrected) drive amplitude to have a simple real-valued Gaussian envelope:
\begin{equation} 
	\kappa (t) = \kappa_0 \exp\left[-\frac{t^2}{t_0^2}\right], 
	\label{eq:SCcontrol}
\end{equation}
with $\kappa_0$ denoting the maximum control amplitude and $t_0$ the effective width of the pulse.
We will consider pulses that turn on at $t = t_{\ui}$ and turn off at $t = t_{\uf}$.

In this problem, the qubit levels are $\ket{0}$ and $\ket{1}$, hence transitions to $\ket{2}$
represent leakage.  We thus write $\hH (t) = \hH_0(t) + \hat{V}(t)$ with 
\begin{equation} 
	\begin{aligned} 
		\hH_0 (t) &= \frac{\delta}{2} \left(-\ketbra{0}{0} +
		\ketbra{1}{1}\right) + \left(\frac{3}{2}\delta + \Delta\right) \ketbra{2}{2} \\
		&\phantom{=} 
		+ \left( \kappa (t) \ketbra{0}{1} + \mm{H.c.} \right) 
	\end{aligned}
	\label{eq:H0drag} 
\end{equation}
and 
\begin{equation} 
	\hV (t) = \lambda \left( \kappa (t) \ketbra{1}{2} + \mm{H.c.} \right).
	\label{eq:Vdrag}
\end{equation}
Note that at a formal level, the parameter $\lambda$ plays the role of $\epsilon$ in Eq.~\eqref{eq:Hgen}.

When the driving field is on resonance with the qubit transition frequency, i.e.~$\delta = 0$, the
unitary evolution operator describing evolution under $\hH_0 (t)$ (starting at a time $t=t_\ui$) is 
\begin{equation} 
	\begin{aligned} 
		\hU_0 (t) &= \cos[\varphi (t)]\left(\ketbra{0}{0} +
		\ketbra{1}{1}\right) - i \sin[\varphi (t)]\left(\ketbra{0}{1} + \ketbra{1}{0}\right) \\ 
		&\phantom{=}+ \exp[-i \Delta (t-t_\ui)] \ketbra{2}{2}, 
	\end{aligned}
	\label{eq:U0drag} 
\end{equation}
with $\varphi (t) = \int_{t_\ui}^t \ud t_1 \kappa (t_1) = \sqrt{\pi} \kappa_0 t_0 [\mm{erf}(t/t_0) -
\mm{erf}(t_\ui/t_0)]/2$. Here, $\mm{erf}(x)$ denotes the error function.  

To implement the Magnus approach, we first write $\hV(t)$ in the interaction picture defined by
$\hU_0 (t)$: 
\begin{equation}
	\begin{aligned} 
		\hV_\uI (t) &= \lambda \left[e^{-i \Delta (t-t_\ui)} \kappa (t)\times \right.\\
		&\phantom{=}
		\left. \left(i \sin[\varphi (t)] \ketbra{0}{2} + \cos[\varphi (t)]
	\ketbra{1}{2}\right) + \mm{H.c.} \vphantom{e^{-i \Delta (t-t_\ui)}}\right].
	\end{aligned} 
	\label{eq:VdragI} 
\end{equation}
We stress again that our choice of interaction picture is very different than that used in DRAG: our
interaction picture is determined by the ideal time-dependent qubit Hamiltonian (including the
driving field on the $\ketbra{0}{1}$ transition).  As such, the desirable qubit drive is not treated
perturbatively. In contrast, DRAG effectively treats all time-dependent terms in
Eq.~\eqref{eq:H0drag} as a perturbation.

The leading order Magnus control correction $\hW_1 (t)$ can now be conveniently found by using the
\textit{derivative-based control} approach [Eq.~\eqref{eq:Hctrl1}]; the second order correction
$\hW_2 (t)$ follows from Eq.~\eqref{eq:Hctrl2}. We find
\begin{equation} 
	\hW_1 (t) = i \frac{\lambda}{\Delta} \dot{\kappa} (t) \ketbra{1}{2} -
	\frac{\lambda}{\Delta}\kappa^2 (t) \ketbra{0}{2} + \mm{H.c.}  
	\label{eq:Hctrl1dragIdeal}
\end{equation}
and
\begin{equation} 
	\begin{aligned} \hW_2 (t) &= \frac{\lambda^2}{\Delta} \kappa^2 (t)
		\left(\ketbra{1}{1}-\ketbra{2}{2}\right) \\ &\phantom{=} -\left(\frac{\lambda^2}{2
		\Delta^2} \kappa^3 (t) \ketbra{0}{1} + \mm{H.c.}\right).  
	\end{aligned}
	\label{eq:Hctrl2dragIdeal} 
\end{equation}
As discussed already, the $\hW_1(t)$ correction generically helps suppress pure leakage transitions
that cause the final qubit state to occupy the $\ket{2}$ level, whereas the $\hW_2(t)$ correction
suppresses phase errors occurring due to virtual transitions to $\ket{2}$.  We have assumed $t_\ui
\gg - 1/\kappa_0$ and $t_\uf \gg 1/\kappa_0$ such that $\kappa (t_\ui) = \kappa(t_\uf) \simeq 0$.

While the above control corrections would ideally suppress the leading errors arising from leakage
in our gate, they are not fully implementable in a typical experimental setup.  In that case, the
only available controls are the (complex) pulse envelope $\kappa(t)$ and drive detuning $\delta(t)$
in Eq.~\eqref{eq:H0drag}.  The control corrections we find differ from those predicted in standard
first order DRAG theory, though DRAG encounters a similar issue: to fully implement the corrections
required (c.f.~Eq.(8) of Ref.~\cite{motzoi2009}), one would e.g.~need to be able to directly control
the $\ketbra{2}{0}$ transition, something not possible in standard setups.

In DRAG, one usually just truncates the unattainable terms in the control Hamiltonian.  However, as
already discussed in Sec.~\ref{sec:imperfectctrls}, simply truncating the ideal controls is not the
best strategy for dealing with unachievable control corrections.  Instead, one should tweak the form
of the realizable parts of $\hW_1(t)$ and $\hW_2(t)$ to reduce the error; various strategies for
doing this were presented in Sec.~\ref{sec:imperfectctrls}.  This provides an alternate and rigorous
explanation for the recent experimental result of  Chen \emph{et al.}~in Ref.~\cite{chen2016}.  In
that work, the authors effectively treated the amplitudes of the two kinds of DRAG corrections
(quadrature control and detuning control) as free parameters.  They demonstrated that by optimizing
over their values, gate fidelities could be improved above what would be obtained by directly using
the leading-order DRAG correction  \footnote{A constant detuning is used to compensate for phase
errors. This allows the authors to implement the correction as a phase shift on
\unexpanded{$\hW_1^{\mm{ctrl}} (t)$}. The constant detuning control corresponds to choose
\unexpanded{$\hW_2 = -\int_{t_\ui}^{t_\uf} \ud t_1 \hW_2 (t) (t_1) /(t_\uf -
t_\ui)$}.}.  In our approach, the need for such an optimization emerges naturally from the fact that
one cannot implement all the terms in the control corrections $\hW_1(t)$ and $\hW_2(t)$.  	

In what follows, we given an explicit demonstration of how to use the \textit{iterative approach} of
Sec.~\ref{sec:imperfectctrls} to minimize errors arising from an imperfect implementation of the
Magnus-based control Hamiltonians.  We start by splitting the $\hW_1(t)$ correction given in
Eq.~\eqref{eq:Hctrl1dragIdeal} into its realizable part $\hW_{1}^{\mm{ctrl}}(t)$ and its remaining,
unrealizable part $\hW_{1}^{\mm{err}}(t)$. Writing these terms in the interaction picture, we have
\begin{equation}
	\begin{aligned} 
		\hW_{1,\uI}^{\mm{ctrl}} (t) &= -\lambda \left[e^{-i \Delta (t-t_\ui)}
		\frac{\dot{\kappa}(t)}{\Delta} \times \right.\\ 
		&\phantom{={}}
		\left. \left(\sin[\varphi(t)] \ketbra{0}{2} - i \cos[\varphi (t)]
	\ketbra{1}{2}\right) + \mm{H.c.} \vphantom{e^{-i \Delta (t-t_\ui)}} \right] 
	\end{aligned}
	\label{eq:Hctr1dragctrlI}
\end{equation}
and 
\begin{equation} 
	\begin{aligned} 
		\hW_{1,\uI}^{\mm{err}} (t) &= -\lambda \left[e^{-i \Delta (t-t_\ui)}
		\frac{\kappa^2 (t)}{\Delta}\times \right. \\ 
		&\phantom{={}} 
		\left. \left(\cos[\varphi (t)] \ketbra{0}{2} + i
		\sin[\varphi(t)]\ketbra{1}{2}\right) + \mm{H.c.} \vphantom{e^{-i \Delta
		(t-t_\ui)}}\right].
	\end{aligned}
	\label{eq:Hctr1dragerrI} 
\end{equation}

Iterating the integration-by-parts procedure described in Sec.~\ref{sec:imperfectctrls_it} two times
shows that we can capture some of the effects of $\hW_{1\uI}^{\mm{err}} (t)$ by modifying the form
of the attainable controls:  $\hW_{1}^{\mm{ctrl}} (t) \rightarrow \hW_{1}^{\mm{ctrl},B} (t)$.  We
find:
\begin{equation}
	\begin{aligned}
	\hW_{1}^{\mm{ctrl},B} (t) &= \lambda\left\{\left[i \left(\frac{\dot{\kappa}(t)}{\Delta} +
	\frac{2}{3}\frac{\rd_t \kappa^3 (t)}{\Delta^3}\right) - \frac{\kappa^3
	(t)}{\Delta^2}\right]\ketbra{1}{2} \right. \\ 
	&\phantom{={}}
	\left . + \mm{H.c.} \vphantom{\left[i \left(\frac{\dot{\kappa}(t)}{\Delta}\right)\right]} \right\}  
	\end{aligned}
	\label{eq:dragMagnusctrl1tmp} 
\end{equation}
Note that this control is not written in the interaction picture, but in the same frame as the
starting Hamiltonian of Eq.~\eqref{eq:H0drag}.  We see that this control Hamiltonian only requires
one to address the $\ket{1} \leftrightarrow \ket{2}$ transition, and not the  $\ket{0}
\leftrightarrow \ket{2}$.  However, in an experiment, one cannot drive the $\ket{1} \leftrightarrow
\ket{2}$ transition independently of the $\ket{0} \leftrightarrow \ket{1}$ transition.  Hence, the
form of the control Hamiltonian in Eq.~\eqref{eq:dragMagnusctrl1tmp} that is actually implementable
is:
\begin{equation} 
	\begin{aligned} 
		\hW_{1}^{\mm{ctrl},C} (t)) &= \left\{\left[i
		\left(\frac{\dot{\kappa}(t)}{\Delta} + \frac{2}{3}\frac{\rd_t \kappa^3
		(t)}{\Delta^3}\right) - \frac{\kappa^3 (t)}{\Delta^2}\right]\times \right.\\ 
		&\phantom{={}}
		\left. \left(\ketbra{0}{1} + \lambda \ketbra{1}{2}\right) + \mm{H.c.}
		\vphantom{\left[i \left(\frac{\dot{\kappa}(t)}{\Delta} + \frac{2}{3}\frac{\rd_t \kappa^3
		(t)}{\Delta^3}\right) - \frac{\kappa^3 (t)}{\Delta^2}\right]} \right\}.
	\end{aligned}
	\label{eq:dragMagnusctrl1o} 
\end{equation}

While this control will help suppress leakage errors, the extra unwanted drive addressing the
$\ket{0} \leftrightarrow \ket{1}$ qubit transition will cause errors in our gate (phase errors).
One thus needs to address this problem before proceeding further; this can be done using our general
framework.  The goal is to ensure that to first order in the Magnus expansion, the error progator
(including our control correction) does not cause any extra evolution in the qubit subspace.  To
ensure this, we will add purely diagonal terms to our control correction: $\hW_{1}^{\mm{ctrl},C} (t)
\rightarrow \hW_{1}^{\mm{ctrl},C} (t) + \hW_{1}^{\mm{diag}} (t)$.  The relevant condition becomes:
\begin{equation} 
	\int_{t_\ui}^{t_\uf} \ud t_1 \hP_Q [\hV_\uI (t_1) +\hW_{1,\uI}^{\mm{ctrl},C} (t)
	(t_1)] \hP_Q + \hW_{1,\uI}^{\mm{diag}} (t_1) = \mathbf{0}.  
	\label{eq:DRAGNoPhaseError}
\end{equation}	

The needed diagonal control Hamiltonian $\hW_{1}^{\mm{diag}} (t)$ can be found using the
\textit{derivative-based control} protocol, yielding:
\begin{equation} 
	\hW_{1}^{\mm{diag}} (t) = 2 \frac{\kappa^2 (t)}{\Delta}\left( 1 +
	\frac{2}{3}\frac{\kappa^2 (t)}{\Delta^2}\right) \left(\ketbra{0}{0} - \ketbra{1}{1} -
	3\ketbra{2}{2}\right).  
	\label{eq:dragMagnusctrl1d} 
\end{equation}
Note that this correction corresponds to a time-dependent variation of the detuning $\delta(t)$ of
the drive pulse, c.f.~Eq.~\eqref{eq:H0drag}.  The term in this control correction $\propto
\ketbra{2}{2}$ is not strictly needed to satisfy Eq.~\eqref{eq:DRAGNoPhaseError}, but was added so
that the full correction is experimentally implementable, i.e.~has the form of a simple detuning
modification. 

\begin{figure}[t!] 
	\includegraphics[width=\columnwidth]{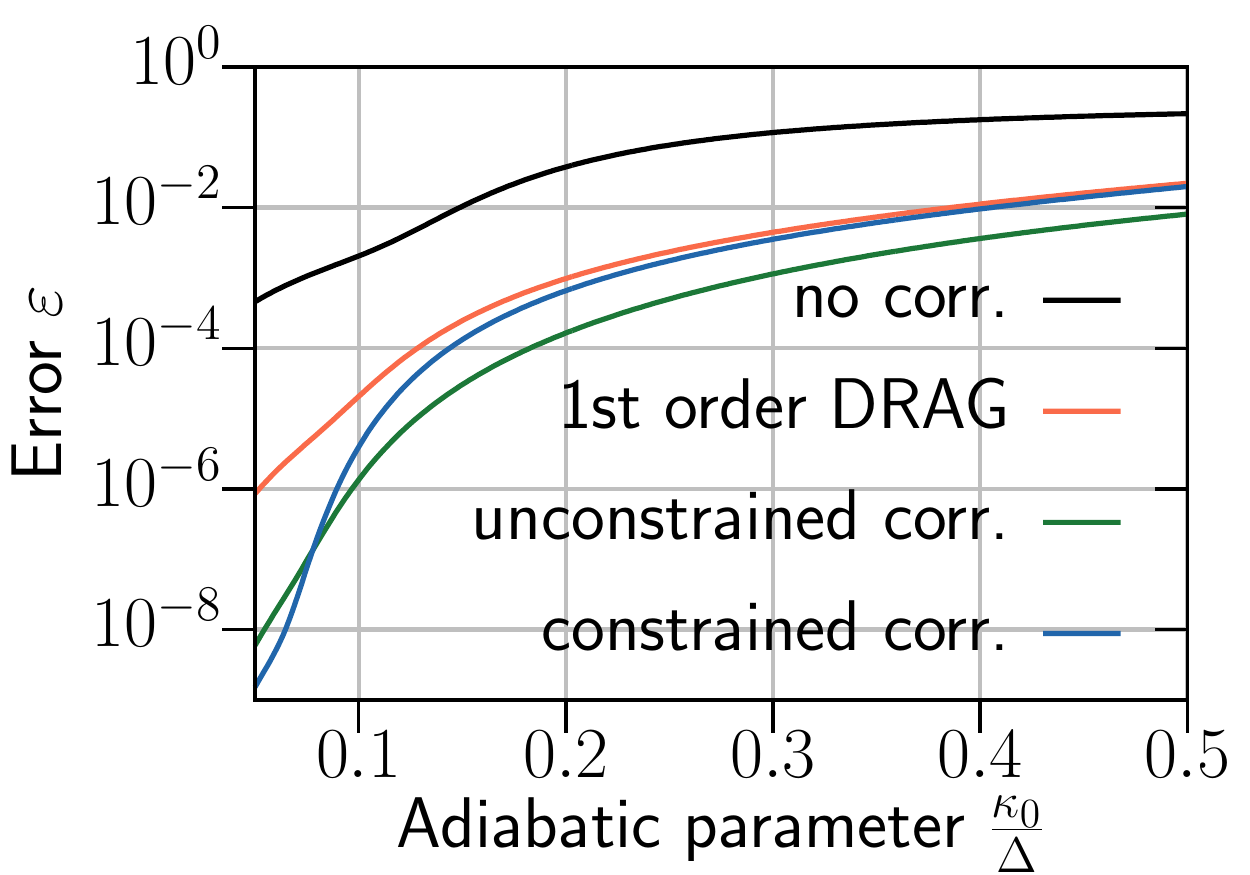} 
	\caption{(Color online). Fidelity error in the implementation of a Hadamard gate in a
		superconducting qubit having an extra leakage level
		c.f.~Eq.~(\ref{eq:DRAGHamiltonian}).  We plot the error in the state-averaged
		infidelity as a function of the adiabatic parameter $\kappa_0/\Delta$. We chose
		$\lambda = \sqrt{2}$, $-t_\ui = t_\uf = 3/\kappa_0$ and $t_0 = \sqrt{\pi}/(4
		\kappa_0)$.  Curves correspond to different choices of corrections: uncorrected
		Gaussian control pulse (black), first-order DRAG corrected control as given in
		Eq.~(9) of Ref.~\cite{motzoi2009} (orange), unconstrained Magnus-based control
		sequence [see Eqs.~\eqref{eq:Hctrl1dragIdeal} and \eqref{eq:Hctrl2dragIdeal}]
		(green), constrained Magnus-based control sequence [see
		Eqs.~\eqref{eq:dragMagnusctrl1} and \eqref{eq:dragMagnusctrl2}] (blue).} 
	\label{fig:drag} 
\end{figure}

We can now summarize and give the final forms of the optimized, experimentally-implementable first
order control Hamiltonian.  It takes the form
\begin{equation} 
	\hW_1^{\mm{eff}} (t) = \hW_{1}^{\mm{diag}} (t) + \hW_1^{\mm{ctrl},C} (t).
	\label{eq:dragMagnusctrl1} 
\end{equation}
where the two terms on the RHS are given by Eqs.~\eqref{eq:dragMagnusctrl1o} and
\eqref{eq:dragMagnusctrl1d}.  Given the form of the first order control, we can immediately use
Eq.~\eqref{eq:Hctrl2} to find the second order control Hamiltonian: 
\begin{equation} 
	\begin{aligned} 
		\hW_2^{\mm{eff}} (t) &= -\frac{1}{2}\frac{\kappa^2
		(t)}{\Delta}\left[\lambda^2 + \frac{\kappa^2 (t)}{\Delta^2}(2-\lambda^2) +
		\frac{4}{3}\frac{\kappa^4 (t)}{\Delta^4}\right]\times \\ 
		&\phantom{={}}
		\left(\ketbra{0}{0} - \ketbra{1}{1} - 3\ketbra{2}{2}\right).  
	\end{aligned}
	\label{eq:dragMagnusctrl2} 
\end{equation}
Here, we have only kept the implementable part of $\hW_2 (t)$.  As a result, it also takes the form
of a time-dependent detuning $\delta(t)$ of the pulse. Using a similar notation to that adopted in
Ref.~\cite{motzoi2009}, and denoting the corrected version of the pulse envelope by
$\tilde{\kappa}(t)$,  the final control sequence is given by 
\begin{equation} 
	\begin{aligned} 
		\tilde{\kappa}_x (t) &= \mm{Re}[\tilde{\kappa} (t)] = \kappa(t) -
		\frac{\kappa^3 (t)}{\Delta^2}, \\ 
		\tilde{\kappa}_y (t) &=  \mm{Im}[\tilde{\kappa}
		(t)] = \frac{\dot{\kappa} (t)}{\Delta} + \frac{2}{3} \frac{\rd_t \kappa^3
		(t)}{\Delta^3}, \\ 
		\delta (t) &= \frac{\kappa^2 (t)}{ \Delta}
		\left(2-\frac{1}{2}\lambda^2\right) +
		\frac{\kappa^4}{\Delta^3}\left(\frac{1}{3}-\lambda^2\right) - \frac{2}{3}
		\frac{\kappa^6 (t)}{\Delta^5}.  
	\end{aligned} 
	\label{eq:dragequivnot} 
\end{equation}

In Fig.~\ref{fig:drag}, we present a comparison of the state-averaged infidelity $\bar{\varepsilon}
= 1 -\bar{F}$, for a Hadamard gate as a function of $\kappa_0 / \Delta$. For simplicity we choose
the Hadamard gate to be $\hU_{\mm{H}} = (\ketbra{0}{0} - i \ketbra{0}{1} - i\ketbra{1}{0} +
\ketbra{1}{1})/\sqrt{2}$.  Without any corrections applied, the gate is achieved using a Gaussian
envelope $\kappa(t)$ of the form in Eq.~\eqref{eq:SCcontrol}, with $t_0 = \sqrt{\pi}/(4 \kappa_0)$;
the total gate time $t_\uf - t_\ui$ is also chosen so that $\kappa(t_\ui) = \kappa(t_\uf) \simeq 0$.
We see that, as expected, the ideal unconstrained Magnus corrections $\hW_1^{\mm{ideal}} (t) +
\hW_2^{\mm{ideal}} (t)$  yield the best fidelity improvements over the uncorrected pulse for the
majority of parameters. For a target infidelity of $\bar{\varepsilon} = 10^{-3}$, these corrections
allow one to reduce the gate time by a factor of $\sim 4$. As shown, this ideal Magnus approach also
yields significant improvements over the standard first order DRAG correction (including the first
order detuning control).

In contrast, if we consider the experimentally-constrained Magnus correction defined by
Eqs.~\eqref{eq:dragMagnusctrl1} and \eqref{eq:dragMagnusctrl2}, we only find significant
improvements over first order DRAG in the quasi-adiabatic regime where $\kappa_0 / \Delta$ is small.
This is also not surprising given the underpinnings of the iterative approach used to derive these
controls: it is only in the limit of a slow gate (compared to $\Delta$) that this approach allows
one to effectively mimic the full Magnus correction using only the experimentally-attainable control
fields.

Fig.~\ref{fig:drag} also reveals that in the extreme limit $\kappa_0 / \Delta \to 0$, the
experimentally-constrained Magnus correction (blue curve) gives a slight advantage over the
unconstrained Magnus correction (green curve).  This might first seem surprising, as the ideal
unconstrained correction eliminates the leading order fidelity error (order $\epsilon^2 \propto
\lambda^2$), whereas the constrained correction only partially cancels this.  While this is true,
the integration-by-parts approach used to derive the constrained correction changes the scaling of
the error in $1 / \Delta$.  This results in a lower overall error in the case where $\kappa_0 /
\Delta \to 0$ while $\lambda$ stays fixed.

\subsection{Multiple-crossings Landau-Zener problem}
\label{sec:mclz}

We now apply our Magnus-based algorithm to a ubiquitous adiabatic passage protocol where there is no
existing good method for fighting leakage:  the so-called multiple-crossings Landau-Zener (MLZ)
model~\cite{demkov1995,usuki1997}.  This is a generalization of the well-known
Landau-Zener-Stückelberg-Majorana (LZ)
model~\cite{landau1932,zener1932,stuckelberg1932,majorana1932} where in addition to the two
desirable ``qubit'' states, several unwanted energy levels are present.  Similar to the LZ model,
the MLZ model describes a variety of important physical systems and control problems.  This includes
state transfer between the electronic spin of a nitrogen vacancy center (NV center) in diamond and
the nuclear spin of nitrogen~\cite{fuchs2011}, and nuclear spin state preparation in self-assembled
quantum dots~\cite{munsch2014}.

The MLZ problem is especially difficult to address, as leakage errors persist even in the limit
where the protocol speed becomes vanishingly small; this is in stark contrast to the STIRAP
state-transfer problem discussed earlier.  In the MLZ problem, couplings to spurious levels cause
the very form of the adiabatic eigenstates of the full system to be corrupted by leakage.  As a
result, a perfect protocol is not realized even if one suppresses all non-adiabatic transitions.   The
MLZ problem cannot be corrected by only using STA techniques, thus illustrating two general
limitations of such approaches: 1) they require exact diagonalization of the instantaneous
Hamiltonian, something that is cumbersome for problems with large Hilbert spaces,  and 2) STA
techniques are only effective if the uncorrected Hamiltonian possesses a set of ``good''
instantaneous (adiabatic) eigenstates whose time-dependent form corresponds to the desired quantum
evolution.

\begin{figure}[t!]
	\includegraphics[width=\columnwidth]{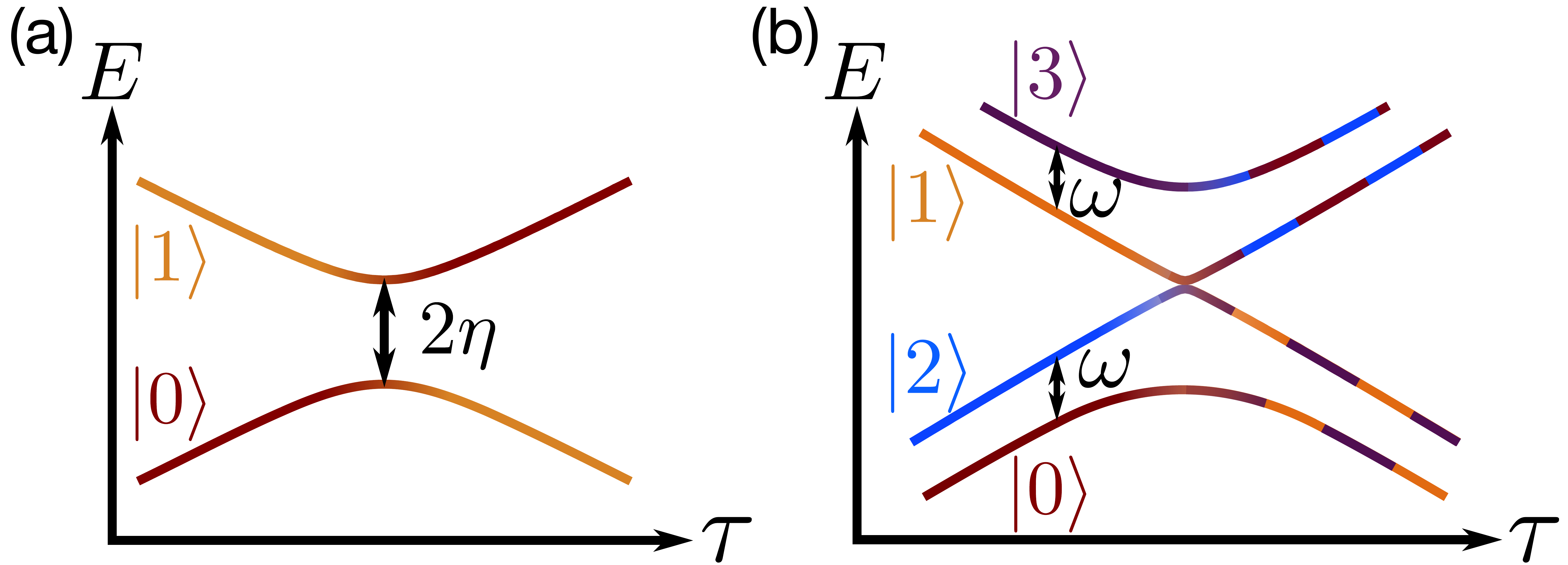}
	\caption{(Color online). (a) Adiabatic energy levels of the
		Landau-Zener-Stückelberg-Majorana model as a function of time.  A purely adiabatic
		evolution realizes an effective NOT gate on the qubit levels $\ket{0},\ket{1}$.
		Colours indicate how the adiabatic eigenstates evolve in the $\ket{0},\ket{1}$
		basis.  (b) A similar plot of adiabatic energy levels versus time for the
		multiple-crossings Landau-Zener model, where in addition to the qubit levels, there
		are two additional leakage levels $\ket{2},\ket{3}$.  Colours again indicate how the
		adiabatic eigenstate wavefunctions evolve in the  $\ket{0},\ket{1},\ket{2},\ket{3}$
		basis.  Because of mixing with the leakage levels, a pure adiabatic evolution no
		longer results in a $\mm{NOT}$ gate of the qubit states. Here, we sketch the
		situation where $\omega_2 = \omega_3 = \omega$, the spurious couplings are
		comparable to $\eta$, and $\omega \lesssim \eta$.}
	\label{fig:multilzsmE} 
\end{figure}

\subsubsection{Landau-Zener state transfer}
\label{sec:lzsm}

We start by reviewing adiabatic passage of a qubit within
the LZ model. The Hamiltonian is~\cite{landau1932,zener1932,stuckelberg1932,majorana1932}
\begin{equation}
	H_{\mm{LZ}} (t) = -\alpha t \hat{\sigma}_z + \lambda \hat{\sigma}_x,
	\label{eq:Hlzsm}
\end{equation}
where $\alpha$ is the energy sweep rate, $\lambda$ the coupling strength between diabatic states,
and $\hat{\sigma}_i$ ($i=x,z$) are Pauli operators.  The $\hat{\sigma}_z$ eigenstates are denoted
$\ket{0}, \ket{1}$, i.e.~$\hat{\sigma}_z = \ketbra{1}{1} - \ketbra{0}{0}$.  The adiabatic dynamics
defined by Eq.~\eqref{eq:Hlzsm} is best understood when introducing the dimensionless time $\tau =
\sqrt{\alpha/\hbar} t$ and dimensionless coupling $\eta = \lambda/\sqrt{\alpha
\hbar}$~\cite{vitanov1996}. Within this framework, Eq.~\eqref{eq:Hlzsm} becomes 
\begin{equation}
	H_{\mm{LZ}} (\tau) = -\tau \hat{\sigma}_z + \eta \hat{\sigma}_x. 
	\label{eq:Hlzsmdimless}
\end{equation}

Consider a protocol where at $\tau = \tau_\ui \ll 0$ the system is initially prepared in its
adiabatic ground, and then as time progresses, sweeps through the avoided level crossing (occurring
at $\tau = 0$).  The protocol ends at a time $\tau = \tau_\uf \gg 0$.  In the infinite-time limit
characterized by $\eta /\abs{\tau_\ui}, \eta /\tau_\uf \ll 1$, the probability that the system
remains in the ground state is given by $P_{\mm{ad}} = 1 - P_{\mm{LZSM}}$, where the non-adiabatic
transition probability $P_{\mm{LZSM}} = \exp[-\pi \eta^2]$.  In the limit $\eta \to \infty$ (or
equivalently $\alpha \to 0$, i.e.~infinitesimally slow sweep rate), the evolution is perfectly
adiabatic, and the system evolves adiabatically between $\ket{0}$ at $\tau_\ui$ to $\ket{1}$ at
$\tau_\uf$, Eq.~\eqref{eq:Hlzsmdimless} [see Fig.~\ref{fig:multilzsmE}~(a)].  Depending on the
physical situation, this can be viewed as having performed a NOT-gate on our qubit, or as having
performed a state transfer operation (as in the NV-system experiment of Ref.~\cite{fuchs2011}).

If there are no other dynamically-relevant energy  levels, one can use STA techniques to find
various control correction Hamiltonians that suppress speed-constraints due to non-adiabatic
transitions.  The simplest control Hamiltonian is the ``transitionless driving'' (TD)
Hamiltonian~\cite{demirplak2003,berry2009}:
\begin{equation}
	\hW_{\mm{TD}} (\tau) = \frac{\eta}{2(\tau^2 + \eta^2)} \hat{\sigma}_y,
	\label{eq:LZtd}
\end{equation}
which was realized experimentally~\cite{bason2012}. Another possibility would be to use the
``superadiabatic transitionless driving'' (SATD) Hamiltonian~\cite{demirplak2008,ibanez2012}:
\begin{equation}
	\begin{aligned}
		\hW_{\mm{SATD}} (\tau) &= \frac{3 \eta^2 \tau}{4 (\tau^2 + \eta^2)^3 +
		\eta^2}\hat{\sigma}_z + \frac{3 \eta \tau^2}{4 (\tau^2 + \eta^2)^3 +
		\eta^2}\hat{\sigma}_x,
	\end{aligned}
	\label{eq:LZsatd}
\end{equation}
whose main advantage compared to $\hW_{\mm{TD}} (\tau)$ is that it does not require a
$\hat{\sigma}_y$ coupling [something which is absent in the original, uncorrected
Hamiltonian of Eq.~\eqref{eq:Hlzsmdimless}].  We stress that both of these corrections allow a perfect state transfer
from $\ket{0}$ to $\ket{1}$.

\subsubsection{Addition of spurious coupled levels}
\label{sec:mclzprob}

\begin{figure}[t!]
	\includegraphics[width=\columnwidth]{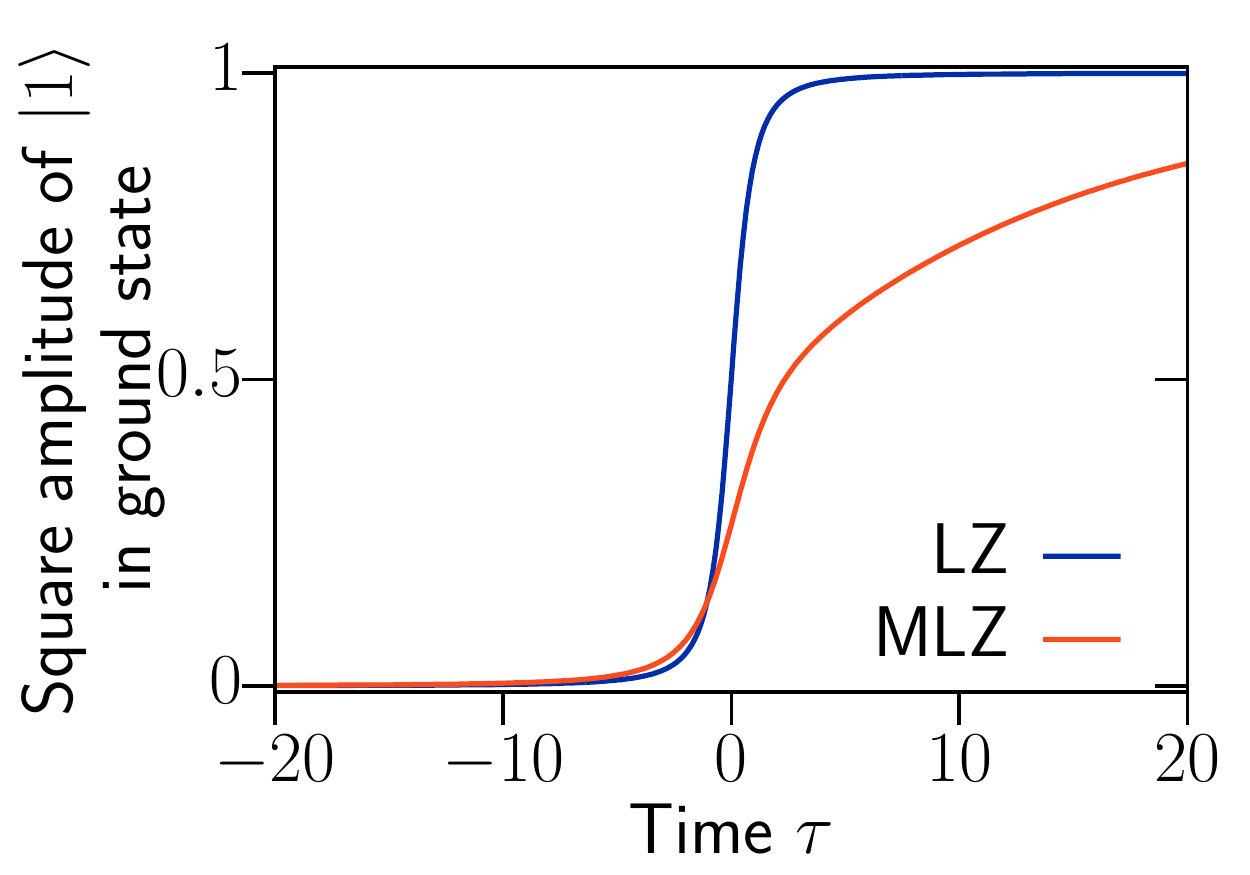}
	\caption{(Color online). Modulus square of the amplitude coefficient of $\ket{1}$
	for the ground state of the Landau-Zener and MLZ model.
	We plot the coefficient as a function of time $\tau$ for $\eta = 1$ and $\omega_2 = \omega_3
	= 0.1$. For the MLZ model, the value is substantially different from $1$ for large values of
	$\tau$. This indicates that a pure adiabatic evolution will not result in a $\mm{NOT}$ gate of
	the qubit states.} 
	\label{fig:GSLZ} 
\end{figure}

Now consider the more general situation described by the multiple-crossings
Landau-Zener~\cite{demkov1995,usuki1997} (MLZ) model, where additional undesirable levels are
coupled to the desired qubit levels [see Fig.~\ref{fig:multilzsmE}~(b)]. For simplicity, we consider
a case where there are two such spurious levels $\ket{2}$ and $\ket{3}$, with the diabatic energy of
$\ket{2}$ ($\ket{3}$) increasing (decreasing) linearly in time.  The Hamiltonian of the system can
be written as
\begin{equation}
	\hH_{\mm{mc}} (\tau) = \hH_{\mm{LZ}} (\tau) + \hH_{\mm{LZ},\mm{aux}} (\tau) + \hV.
	\label{eq:multilzsm}
\end{equation}
Here, $\hH_{\mm{LZ}} (\tau)$ is given again by Eq.~\eqref{eq:Hlzsmdimless} and describes the
desirable dynamics of the qubit levels $\ket{0},\ket{1}$.  In contrast,  $\hH_{\mm{LZ},\mm{aux}}
(\tau)$ describes the dynamics within the leakage subspace, and $\hV$ describes the undesirable
coupling between qubit and leakage levels.  We have 
\begin{equation}
	\begin{aligned}
	\hH_{\mm{LZ}} (\tau) &= \tau \left(\ketbra{0}{0} - \ketbra{1}{1} \right) + \eta
	\left(\ketbra{0}{1} + \ketbra{1}{0} \right), \\
	\hH_{\mm{LZ},\mm{aux}} (\tau) &= 
		\left(\tau + \omega_2 \right) \ketbra{2}{2} - 
		\left(\tau - \omega_3 \right)\ketbra{3}{3} \\
	&\phantom{={}}
	+ \eta_{23} \left(\ketbra{2}{3} + \ketbra{2}{3} \right),\\
	\hV  &= \eta_{12}  (\ketbra{1}{2} + \ketbra{2}{1}) +
	\eta_{03} (\ketbra{0}{3} + \ketbra{3}{0}),
	\end{aligned}
	\label{eq:multilzsndef}
\end{equation}
We have assumed for simplicity that the spurious levels have an energy detuning rate that is the
same as the qubit levels. $\omega_2$ denotes the constant-in-time separation between the diabatic
energies of the upward-moving qubit level $\ket{0}$ and upward-moving leakage level $\ket{2}$.
Similarly, $\omega_3$ denotes the separation between the diabatic energies of the $\ket{1}$ and
$\ket{3}$ states.  Finally, the $\eta_{ij}$ denote couplings involving the leakage levels (including
couplings to the qubit states).  We only retain couplings between levels whose diabatic energies can
cross.  

We can immediately see from Eq.~\eqref{eq:multilzsndef} that a simple adiabatic evolution will no
longer allow us to perform the desired gate on the qubit levels (i.e.~an effective NOT-gate on
the qubit levels $\ket{0},\ket{1}$). In the ideal case, the adiabatic ground state of the full
Hamiltonian should reduce to $\ket{0}$ at the start of the protocol and to  $\ket{1}$ at the end of
the protocol. The presence of the spurious levels will prevent this from being the case. For $\eta
/\abs{\tau_\ui} \ll 1$ but finite, the qubit state $\ket{0}$ and leakage state $\ket{2}$ will be weakly coupled
to one another (via a virtual process involving the $\ket{3}$ state), leading the adiabatic ground
state to have an appreciable and unwanted $\ket{2}$ component.  Similarly, for $\eta/\tau_\uf \ll1$ but finite,
the ground state will be an admixture of the qubit state $\ket{1}$ and leakage state $\ket{3}$ (see
Fig.~\ref{fig:multilzsmE}~(b) and Fig.~\ref{fig:GSLZ}).

Before proceeding, we stress that several experimental systems are described exactly by a
Hamiltonian of the form of Eq.~\eqref{eq:multilzsm} (where both desirable and undesirable levels
have diabatic energies changing linearly in time).  For example, consider the experiment of
Ref.~\cite{fuchs2011}, where the authors demonstrate the realization of a quantum memory using a NV
center in diamond.  Using adiabatic passage under the linear ramp of a magnetic field, they are able
to transfer the state of the electronic spin of the NV center into the nuclear spin of nitrogen. In
this system, it is the hyperfine interaction that allows one to perform state transfer between the
two states of interest.  However, imperfect magnetic field alignment implies that there are always
unwanted coupling to other, unwanted electronic and nuclear spin states.  This problem can be mapped
exactly onto the multiple-crossings Landau-Zener problem described here.
 
\subsubsection{Magnus-based control corrections} 
\label{sec:mclzcorrs}

To derive control corrections for our MLZ system using the Magnus-based approach, we will take a
two-step approch:
\begin{itemize}
	\item First, ignore the presence and coupling to the leakage levels, and use the SATD
		approach to cancel errors arising from non-adiabatic transitions between the
		adiabatic eigenstates of the qubit Hamiltonian $\hH_{\rm LZ}(\tau)$.  This involves
		using the control-correction Hamiltonian  $\hW_{\mm{SATD}} (\tau)$ given in
		Eq.~\eqref{eq:LZsatd}.
	
	\item Next, include the coupling to the leakage levels.  Use a Magnus-based correction to
		cancel the effects of $\hV$ in Eq.~\eqref{eq:multilzsm}.  
\end{itemize}

We thus need to find corrections starting with the Hamiltonian 
\begin{equation}
	\hH (\tau) = \hH_{\mm{mc}} (\tau)  + \hW_{\mm{SATD}} (\tau).
	\label{eq:HtotmcEx}
\end{equation}
Note that here, the Magnus-algorithm is only fighting errors arising from the coupling to the
leakage levels; non-adiabatic errors within the qubit subspace are fully addressed using the known
ideal control $\hW_{\mm{SATD}} (\tau)$.  

\begin{figure}[t!]
	\includegraphics[width=\columnwidth]{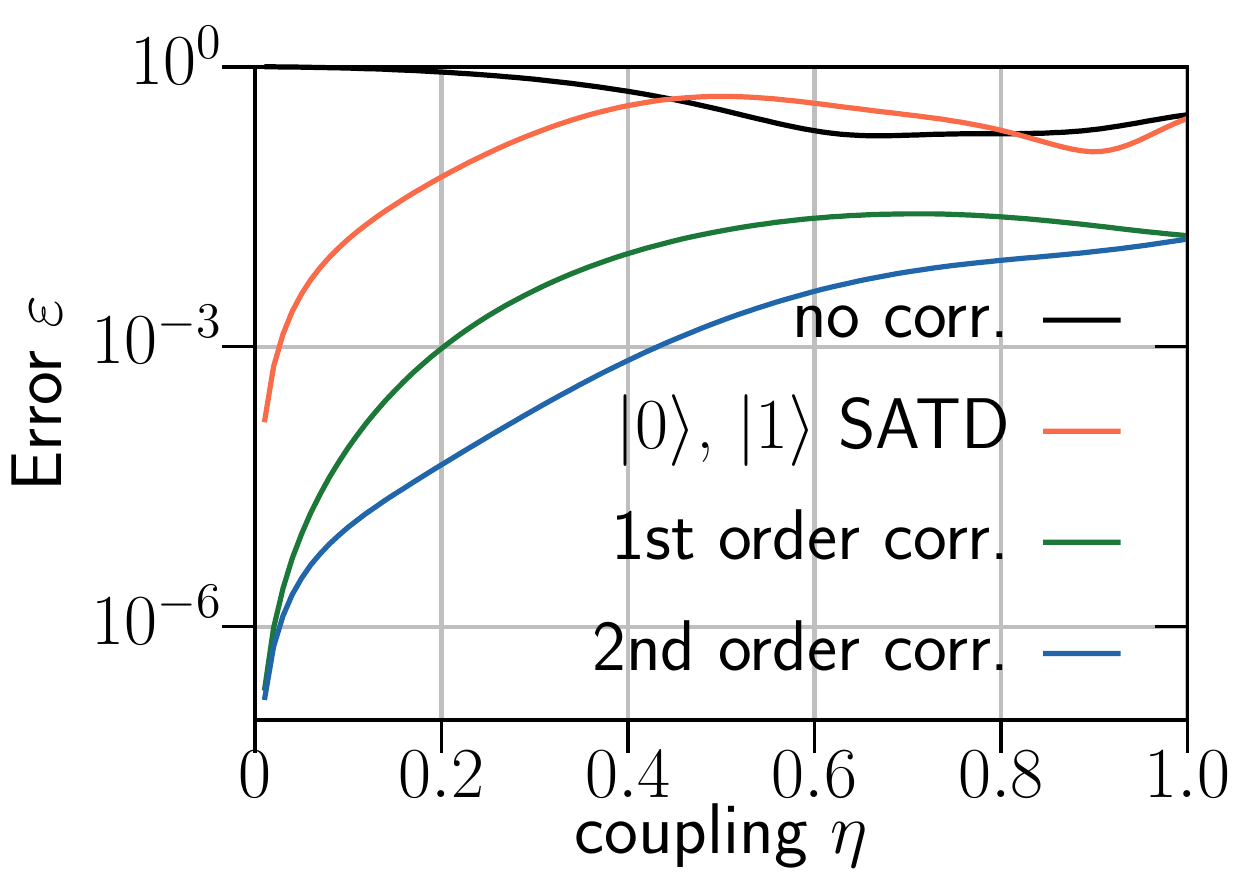}
	\caption{(Color online).  Fidelity error for an attempted $\ket{0} \rightarrow \ket{1}$
	state transfer in the MLZ model with two leakage levels [c.f.~Eq.~\eqref{eq:multilzsm}], as
	a function of the dimensionless coupling $\eta$.  We have taken the couplings involving
	spurious levels to be the same as the coupling $\eta$ between qubit levels, i.e. $\eta_{03}
	= \eta_{12} = \eta_{23} = \eta$, and have set the energy offsets $\omega_2 = \omega_3 =
	0.1$.  Different curves correspond to different possible corrections (as indicated in the
	figure).  Without any correction, small errors are not possible for any choice of $\eta$; in
	contrast, if one includes the Magnus corrections, errors less than $10^{-3}$ are possible
	for a wide range of $\eta$.}
	\label{fig:multilzsm} 
\end{figure}

To apply the Magnus-approach, it is convenient to work in the so-called superadiabatic
frame~\cite{garrido1964,berry1987,berry1990} of the qubit subspace.  One first moves to the
adiabatic frame of the qubit Hamiltonian $\hH_{\mm{LZ}}(\tau)$.  One then finds the unitary that
diagonalizes this adiabatic Hamiltonian at each time.  Using this time-dependent unitary defines the
final superadiabatic frame.  In this frame, the Hamiltonian reads 
\begin{equation}
	\hH_{\mm{SAD}} (\tau) = \hH_{0,\mm{SAD}} (\tau) + \hV_{\mm{SAD}} (\tau),
	\label{eq:multilzsmsad}
\end{equation}
where the subscript ``SAD'' indicates an operator written in the superadiabatic frame. We have 
\begin{equation}
	\begin{aligned}
		\hH_{0,\mm{SAD}} (\tau) = &-\frac{\sqrt{4 (\tau^2 + \eta^2)+\eta^2}}{2 (\tau^2 + \eta^2)}
		\left(\ketbra{\tilde{1}}{\tilde{1}}-\ketbra{\tilde{0}}{\tilde{0}}\right) \\
		&+
		(\tau + \omega_2) \ketbra{2}{2} - (\tau - \omega_3)
		\ketbra{3}{3},
	\end{aligned}
	\label{eq:Hlzsad0}
\end{equation}
with $\ket{\tilde{j}} = \hS^{\dag} (\tau) \ket{j}$ ($j=0,1$). The unitary operator $\hS^{\dag}
(\tau)$ is the change of basis operator that takes one to the superadiabatic frame; it only acts
non-trivially on the computational subspace. The operator $\hS (\tau)$ is readily found since it
corresponds diagonalizing a $2 \times 2$ matrix.

\begin{figure*}[t!]
	\includegraphics[width=2\columnwidth]{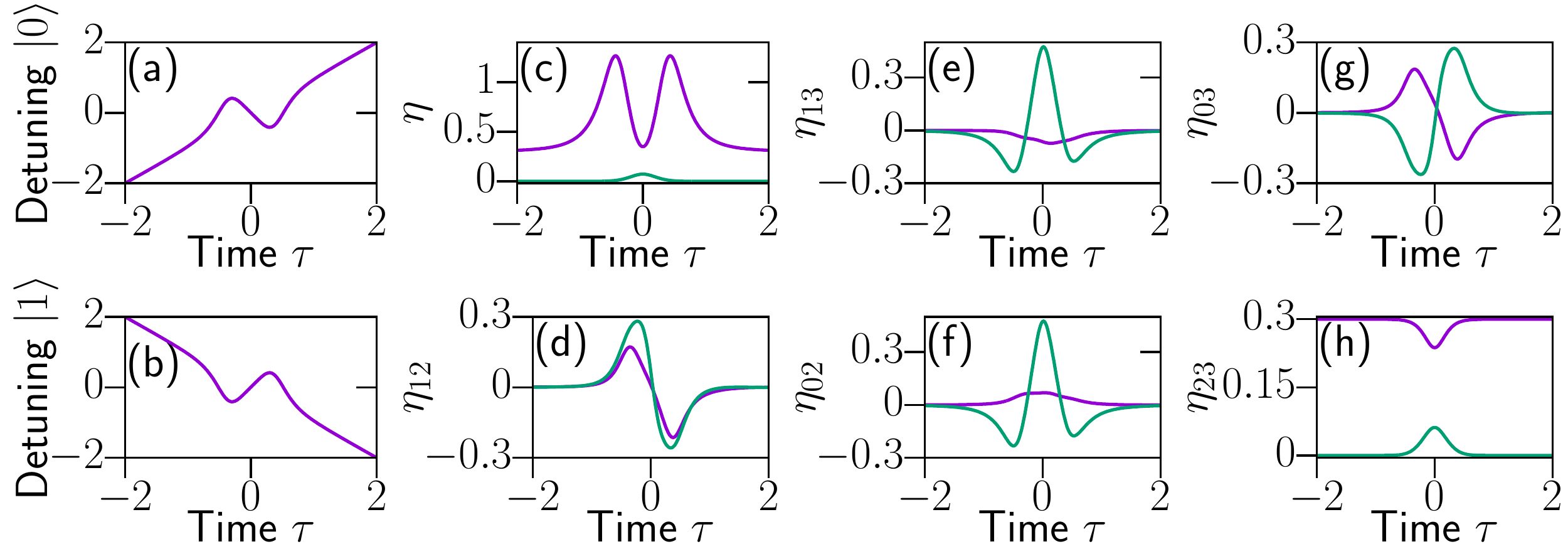}
	\caption{(Color online). Real (purple) and imaginary (green) part of the controls predicted
		by the Magnus-based algorithm for the MLZ model as a function of time $\tau$ for
		$\eta = \eta_{12} = \eta_{03} = \eta_{23}=0.3$ and $\omega_2 = \omega_3 = 0.1$. (a)
		Detuning of $\ket{0}$. (b) Detuning of $\ket{1}$. (c) Coupling $\eta$ between the
		qubit states $\ket{0}$ and $\ket{1}$. (d) Coupling between $\ket{1}$ and the leakage level
		$\ket{2}$. (e) Coupling between $\ket{1}$ and the leakage level $\ket{3}$. (f)
		Coupling between $\ket{0}$ and the leakage level $\ket{2}$. (g) Coupling between
		$\ket{0}$ and the leakage level $\ket{3}$.  Coupling between the leakage levels
		$\ket{2}$ and $\ket{3}$.}
	\label{fig:multilzsmctrls} 
\end{figure*}

The leakage Hamiltonian $\hV$ becomes time-dependent in the superadiabatic frame.  The matrix
elements coupling leakage and qubit levels have the general form
\begin{equation}
	\begin{aligned}
		\sq \hV_{\mm{SAD}} (\tau) 
		&=V_{12} (\tau) \ketbra{\tilde{1}}{2} + V_{13} (\tau)
		\ketbra{\tilde{1}}{3} \\ 
		&\phantom{= {}} + V_{02} (\tau) \ketbra{\tilde{0}}{2} + V_{03} (\tau)
		\ketbra{\tilde{0}}{3} + \mm{H.c.},
	\end{aligned}
	\label{eq:Vleaklzsad}
\end{equation}
Explicit expressions for the matrix elements of $\sq \hV_{\mm{SAD}} (\tau)$ are readily found and
are given in Appendix~\ref{sec:Vleaklzsadel}.  The $\hV_{\mm{SAD}} (\tau)$ operator also couples
leakage levels to one another, the form here is the same as in the original (lab) frame:
\begin{equation}
	P_{N-Q} \hV_{\mm{SAD}} (\tau) P_{N-Q} = \eta_{23} (\ketbra{2}{3} +
	\ketbra{3}{2}).
	\label{eq:Vsubleak}
\end{equation}

Equation~\eqref{eq:multilzsmsad} has of course the general form of the leakage problem addressed in
this paper.  We can then proceed in applying our Magnus-based algorithm to minimize the effects of
the leakage Hamiltonian $\hV_{\mm{SAD}}(\tau)$. As always, the first step consists in finding a
control correction $\hW_1(\tau)$ that averages away the pure leakage interaction described by $\sq
\hV_{\mm{SAD}} (\tau)$. We will focus on a protocol that runs between $\tau = \tau_\ui \ll 0$ and
$\tau = \tau_\uf \gg 0$.  Since $\sq \hV_{\mm{SAD}} (\tau_\ui) \neq \mathbf{0}$ and $\sq
\hV_{\mm{SAD}} (\tau_\uf) \neq \mathbf{0}$, we cannot rely on the \textit{derivative-based control}
to find $\hW_{\mm{SAD},1} (\tau)$. We thus instead use the \textit{generating function approach}
(c.f.~Sec.~\ref{sec:genfct}).  This involves first picking a time-dependent operator $\hR (\tau)$
that is similar in form to the leakage operator $\hV_{\mm{SAD}} (\tau)$, but which vanishes at $\tau
= \tau_\uf$.  We take
\begin{equation}
	\hR (\tau)  = \ell_0 (\tau) \hV_{\mm{part}} (\tau),
	\label{eq:Omega1multilzsm}
\end{equation}
where the superoperator $\ell_0$ takes us into the interaction picture generated by
$\hH_{0,\mm{SAD}}$ [c.f.~Eq.~\eqref{eq:SuperOpL}], and
\begin{equation}
	\begin{aligned}
		\hV_{\mm{part}} (\tau) &= \left(\mm{Re}\left[V_{12} (\tau) \right] \ketbra{\tilde{1}}{2} +
		\mm{Re}\left[V_{13} (\tau) \right] \ketbra{\tilde{1}}{3} \right.\\ 
		&\phantom{= {}} 
		+ i \mm{Im}\left[V_{02} (\tau)\right]\ketbra{\tilde{0}}{2} +
		i  \mm{Im}\left[V_{03} (\tau)\right] \ketbra{\tilde{1}}{3} \\
		&\phantom{= {}}
		\left. + \mm{H.c.}\right).
	\end{aligned}
	\label{eq:Hctrl1lz}
\end{equation}
We have essentially kept the parts of $\hat{V}_{\mm{SAD}}(\tau)$ which vanish for large enough
$\tau_\uf$.  

Next, we pick a control correction so that the first order expansion of the error propagator
$\hOmega_1^{(1)}(\tau)$ is equal to $\hR (\tau)$ (and hence vanishes at large $\tau$).  Using
Eq.~\eqref{eq:Hctrl1Mag1}, we find 
\begin{equation}
	\hW_1 (\tau) = \ell_0^\dag (\tau) \left\{\rd_{\tau} \left[\ell_0 (\tau) \hV_{\mm{part}}
	(\tau)\right] - \sq \hV_{\mm{SAD},\uI} (\tau)\right\},
	\label{eq:altstart}
\end{equation}
where $\sq \hV_{\mm{SAD},\uI} (\tau) = \ell_0 \sq \hV_{\mm{SAD}}(\tau)$ is the full leakage operator
in the interaction picture.  Given the form of the first order control correction Hamiltonian
$\hW_1(\tau)$, the second order correction $\hW_2 (\tau)$ follows immediately from
Eq.~\eqref{eq:Hctrl2}. 

In Fig.~\ref{fig:multilzsm}, we plot the fidelity error $\varepsilon$ versus coupling $\eta$ for a
state transfer protocol where we start in state $\ket{0}$ at $\tau = \tau_\ui$ and evolve under
$\hH_{\mm{mc}} (\tau)$ (plus possible control corrections) until a time $\tau = \tau_\uf$, with the
goal of reaching $\ket{1}$. We consider the extreme situation where all inter-level couplings (both
desirable and undesirable) are the same:  $\eta = \eta_{23} = \eta_{03} = \eta_{12}$.  We also have
fixed $\omega_2 = \omega_3 = 0.1$ and taken $\tau_\ui = -20$ and $\tau_\uf = 20$.  We see that
without any corrections, the evolution generated by $\hH_{\mm{mc}} (\tau)$ (black curve) never
permits an error lower than $10^{-1}$. For fast protocols (small $\eta$) non-adiabatic errors limit
the error, whereas for slow protocols (large $\eta$) the error is limited by the corruption of the
adiabatic eigenstates by the leakage levels.  While applying the qubit-only SATD correction gives
improvements for a narrow range of $\eta$ (orange curve), we see that the Magnus-based corrections
yield better improvements over the range of $\eta$ shown in Fig.~\ref{fig:multilzsm}. In
particular, errors much lower than the $10^{-3}$ level required for quantum error correction are
possible.  

In Fig.~\ref{fig:multilzsmctrls}, we plot the controls predicted by the Magnus-based approach. As
one can observe, the modified protocol requires both control of the detunings [diagonal elements of
$\hH_{\mm{tot},ij}^{(2)} (\tau)$] and couplings. Note also that the control fields are smooth
functions of time.  

Although our choice for $\sq \hOmega_1^{(1)} (t)$ [Eq.~\eqref{eq:Omega1multilzsm}] was not motivated
by any experimental realization, the controls found here could in principle be implemented in a
NV-center in diamond in the manifold spanned by $\{\ket{m_s =-1,m_n = 0}, \ket{m_s =-1,m_n = +1},
\ket{m_s =0 ,m_n = 0}, \ket{m_s =0,m_n = -1}\} \equiv \{\ket{0}, \ket{2}, \ket{3}, \ket{1}\}$. Here
$m_s$ labels the spin projection in the $S=1$ ground state of the NV and $m_n$ labels the nuclear
spin projection of the nitrogen. These controls are possible since the detuning is achieved by
sweeping an external magnetic field~\cite{fuchs2011}, microwaves fields can be used to manipulate
the transition $\ket{m_s =-1,m_n = 0} \leftrightarrow \ket{m_s =0,m_n = 0}$~\cite{jelezko2004}, and
the nuclear spin transitions could be addressed with the method developed in
Ref.~\cite{taminiau2012}.

\section{Conclusion}
\label{sec:conc}

We have presented a systematic perturbative method based on the Magnus expansion that allows one to
construct corrections to time-dependent control Hamiltonians in such a way that leakage errors
during a quantum gate are minimized.  The method is very general, and can be applied to problems
where one does not have full control over the system.  We have discussed how our method connects to
both the well known techniques of DRAG and shortcuts to adiabaticity (STA), emphasize how in certain
cases, it is able to overcome their limitations.  

\section{Acknowledgements}
\label{sec:ackowledgements}

We acknowledge funding from the University of Chicago Quantum Engineering program and the AFOSR MURI
program.  H. R. gratefully acknowledges funding from the Swiss SNF. 

\begin{appendix}

\section{Details of cancellations for second order control}
\label{sec:cancellations}

Let us verify the claim in the main text, that the choice for $\hW_2(t)$ given in
Eq.~\eqref{eq:Hctrl2avg} explicitly cancels errors to order $\mathcal{O}[\pp^2]$.  To do this, we
explicitly compute $\hOmega^{(2)} (t_\uf)$ to second order in $\hV_\uI (t)$. The first term in the
Magnus expansion of  $\hOmega^{(2)}(t)$ is given by 
\begin{equation}
	\begin{aligned}
		\hOmega_1^{(2)} (t) &=  -i \int_0^{t} \ud t_1 \left\{\left[\pp \sq \hV_\uI
		(t_1) + \hW_{1,\uI} (t_1) \right] + \hW_{2,\uI} (t_1) \right. \\
		&\phantom{={}} \left.+ \epsilon\left[\hV_\uI (t_1) - \sq \hV_\uI (t_1)
		\right]\right\}.
	\end{aligned}
	\label{eq:Omega1_2}
\end{equation}i
At $t_\uf$, we have 
\begin{equation}
	\sq \hOmega_1^{(2)} (t_\uf) = -i \int_0^{t_\uf} \ud t_1 \hW_{2,\uI} (t_1),
	\label{eq:Omega1_2_tf}
\end{equation}
where we used Eq.~\eqref{eq:Hctrl1gen}, $\sq^2 \hO (t) = \sq \hO (t)$, and that $\hW_2 (t)$ does not
act on the leakage subspace, $\sq \hW_2 (t) = \hW_2 (t)$. For convenience we have also defined
$\rd_t \hZ (t) = \epsilon[\hV_\uI (t) - \sq \hV_\uI (t)]$. 

The second term in the Magnus expansion of $\hOmega^{(2)}(t)$ is given by
[c.f.~Eq.~\eqref{eq:magnus_series}]:
\begin{equation}
	\hOmega_2^{(2)} (t_\uf) = \frac{1}{2}\int_0^{t_\uf} \ud t_1 \left[\rd_{t_1} \hOmega_1^{(2)}
		(t_1), \hOmega_1^{(2)} (t_1)\right].
\end{equation}
Substituting in the result of Eq.~\eqref{eq:Omega1_2}, this becomes
\begin{equation}
	\begin{aligned}
		\hOmega_2^{(2)} (t) &=\frac{1}{2}\int_0^{t} \ud t_1 \left\{\com{\rd_{t_1} \sq
		\hOmega_1^{(1)} (t_1)}{\sq \hOmega_1^{(1)} (t_1)} \right. \\
		&\phantom{={}} 
		+ \com{\rd_{t_1} \hZ (t_1)}{\hZ (t_1)} 
		+ \com{\rd_{t_1} \hZ(t_1)}{\sq\hOmega_1^{(1)} (t_1)} \\
		&\phantom{={}} \left.
		+ \com{\rd_{t_1} \sq \hOmega_1^{(1)} (t_1)}{\hZ(t_1)} \right\} \\
		&\phantom{={}} 
		+ \mathcal{O}(\epsilon^3) \\
		&= \sq \left[\hOmega_2^{(1)} (t) - \hOmega_2^{(1)} (0)\right] \\
		&\phantom{={}} 
		+ \frac{1}{2} \int_0^{t} \ud t_1 \com{\rd_{t_1} \hZ(t_1)}{\hZ (t_1)} +
		\mathcal{O}(\epsilon^3).
	\end{aligned}
	\label{eq:Omega2_2}
\end{equation}
We then have
\begin{equation}
	\sq \hOmega_2^{(2)} (t_\uf) = \sq \left[\hOmega_2^{(1)} (t_\uf) - \hOmega_2^{(1)}
	(0)\right],
	\label{eq:Omega2_2_tf}
\end{equation}
where once more we have used $\sq^2 \hO (t) = \sq \hO (t)$ and the fact that $\com{\rd_{t}
\hZ(t)}{\hZ (t)}$ is an operator that only acts on the leakage subspace.

Finally, combining the results of Eqs.~\eqref{eq:Omega1_2_tf} and \eqref{eq:Omega2_2_tf}, we get 
\begin{equation}
	\sq \hOmega^{(2)} (t_\uf) = \mathcal{O}\left(\epsilon^3\right),
	\label{eq:proofW2f}
\end{equation}
if we choose 
\begin{equation}
	-i \int_0^{t_\uf} \ud t_1 \hW_{2,\uI} (t_1) = \sq \left[\hOmega_2^{(1)} (t_\uf) -
	\hOmega_2^{(1)} (0)\right].
	\label{eq:proofW2} 
\end{equation}

\section{Additional features of the Magnus-expansion derived controls}
\label{sec:magnusexpansion}

\subsection{Convergence of the series defining the control Hamiltonian}
\label{sec:magnus_conv}

We have shown how to get control Hamiltonians based on the Magnus expansion. There is, however, an
issue we have not addressed; namely the convergence problem. One would naturally expect the method
to yield results whenever the Magnus expansion converges.  In particular, the Magnus expansion for the
modified error propagator $\hU^{(j)}_{\mm{err}}$ (which includes the first $j$ control corrections $\hW_j(t)$) 
is guaranteed to converge if the following condition holds~\cite{casas2007,moan2008}
\begin{equation}
	\int_0^t \ud t_1 \norm{\hV_\uI^{(j)} (t_1)}_2 < \pi,
	\label{eq:Magnusconv}
\end{equation}
with $\norm{\cdot}_2$ denoting the $p=2$-norm.  This condition allows one to make two strong
statements about the Magnus-based algorithm: 1) it is possible to choose $\hW_k (t)$ such that
Eq.~\eqref{eq:Magnusconv} is always fulfilled, 2) since the method is based on perturbation theory
within a suitable interaction picture, there is always a parameter regime where
Eq.~\eqref{eq:Magnusconv} holds.

Opposite to optimal control methods~\cite{khaneja2005,cerfontaine2014}, we expect the Magnus-based
algorithm to yield smooth control sequences. If $\hV (t)$ is a smooth function of time (at least
$C^0$), then it is always possible to find controls that are also smooth. 

\subsection{Impact on the fidelity}
\label{sec:fid}

In this appendix, we show that implementing $\hW_1 (t)$ [see Eq.~\eqref{eq:Hctrl1gen}] predicted by
the Magnus-based algorithm leads to an expression for the fidelity whose leading order correction
scale as $\epsilon^4$. 

First, we consider the state-dependent fidelity for the unitary evolution generated by
$\hH (t) + \hW_1 (t)$. For an initial state $\ket{\psi_0}$ is in $\HHcomp$, we have
\begin{equation}
	\begin{aligned}
		F &= \abs{\bra{\psi_0} \hU_0^\dag (t_\uf) \hU (t_\uf) \ket{\psi_0}}^2 \\
		  &= \abs{\bra{\psi_0} \hP_Q \hU_0^\dag (t_\uf) \hU (t_\uf) \hP_Q \ket{\psi_0}}^2 \\
		  &= \abs{\bra{\psi_0} \hP_Q \hU_{\mm{err}}^{(1)} (t_\uf) \hP_Q \ket{\psi_0}}^2 \\
		  &= \abs{\bra{\psi_0} \hP_Q \exp\left[\sum_{k=1}^\infty \hOmega_k^{(1)}
		     (t_\uf)\right] \hP_Q \ket{\psi_0}}^2,
	\end{aligned}
	\label{eq:statefiddemo1}
\end{equation}
where we have used the fact that $\hU (t_\uf) = \hU_0 (t_\uf) \hU_{\mm{err}}^{(1)} (t_\uf)$ and the
identity  $\hP_Q \ket{\psi_0} = \ket{\psi_0}$.  Expanding the exponential operator in
Eq.~\eqref{eq:statefiddemo1} up to $\epsilon^4$ yields
\begin{equation}
	\begin{aligned}
	&F \simeq \\
	&\abs{\bra{\psi_0} \hP_Q \left\{\mathbbm{1} + \sum_{k=2}^4
	\hOmega_k^{(1)} (t_\uf) + \frac{1}{2} \left[\hOmega_2^{(1)}
	(t_\uf)\right]^2 \right\} \hP_Q \ket{\psi_0}}^2,
	\end{aligned}
	\label{eq:statefiddemo2}
\end{equation}
where, by virtue of our choice of interaction picture [see Eqs.~\eqref{eq:Ugen} and
\eqref{eq:SuperOpL}], we have $\hP_Q \hOmega_1^{(1)} (t) \hP_Q = \mathbf{0}$ for all times.
Furthermore, since $\hOmega^{(1)} (t_\uf) = \sq \hOmega^{(1)} (t_\uf) + \hP_{N-Q} \hOmega^{(1)}
(t_\uf) \hP_{N-Q} = \hP_{N-Q} \hOmega^{(1)} (t_\uf) \hP_{N-Q}$, we have $\hP_Q [\hOmega_1^{(1)}
(t_\uf)]^2
\hP_Q = \mathbf{0}$.

Finally, using the relation $\abs{\bra{\varphi}\hO\ket{\varphi}}^2 = \bra{\varphi}\hO\ket{\varphi}
\bra{\varphi}\hO\ket{\varphi}^\ast =
\bra{\varphi}\hO\ket{\varphi}\bra{\varphi}\hO^\dag\ket{\varphi}$ and  $\hOmega_k^\dag (t) = -
\hOmega_k (t)$, we find 
\begin{equation}
	F = 1 - \mathcal{O}(\epsilon^4).
	\label{eq:statefiddemo3} 
\end{equation}

The result for the averaged-state fidelity follows from Eq.~\eqref{eq:statefiddemo3}, since the
state-averaged fidelity is formally defined as~\cite{pedersen2007}
\begin{equation}
	\bar{F} = \frac{1}{Q} \sum_{\ket{\psi_0}} F.
	\label{eq:avgfiddefsum}
\end{equation}
A similar calculation for $\hH (t) + \hW_1 (t) + \hW_2 (t)$ shows that the leading order correction to
the fidelity, both state-dependent and state-averaged, scales like $\mathcal{O}(\epsilon^6)$.

\section{Relation between Magnus-based controls and \texorpdfstring{$\hW_{\mm{SATD}} (t)$}{WSATD}
for STIRAP with constant gap} 
\label{sec:compsatd}

In the interaction picture defined by $\hH_0$ in Eq.~\eqref{eq:Hadlambda}, the SATD control
Hamiltonian for the constant gap STIRAP reads
\begin{equation}
	\begin{aligned}
		\hW_{\mm{SATD},\uI} (t) &= \frac{\ddot{\theta} (t)}{\sqrt{2} G_0} \left[1 +
		\frac{\dot{\theta}^2 (t)}{G_0^2}\right]^{-1} \times \\
		&\phantom{={}} \left( e^{i G_0 t} \ketbra{\ud}{\ub_-} - e^{-i G_0 t}
		\ketbra{\ud}{\ub_+} + H.c.\right) \\
		&=\frac{\ddot{\theta} (t)}{\sqrt{2} G_0} \sum_{k=0}^\infty (-1)^k
		\left[\frac{\dot{\theta}^2(t)}{G_0^2}\right]^{2k}\times \\
		&\phantom{={}} \left( e^{i G_0 t} \ketbra{\ud}{\ub_-} - e^{-i G_0 t}
		\ketbra{\ud}{\ub_+} + \mm{H.c.}\right).
	\end{aligned}
	\label{eq:satdintconstgap}
\end{equation}
As the second equality suggests, we can perform an expansion in powers of $\dot{\theta} (t)
/ G_0$. The $k=0$ term corresponds to $\hW_{1,\uI} (t)$. However, Eq.~\eqref{eq:satdintconstgap}
cannot yield a correction resembling $\hW_{2,\uI} (t)$. 

\section{Approximate second-order optimal control \texorpdfstring{$\hW_{2,\mm{opt},\mm{ad}}
(t)$}{W2opt} for STIRAP with a constant gap} 
\label{sec:optctrlSTIRAP}

In this section, we show how we obtained the approximate second order optimal control
$\hW_{2,\mm{opt},\mm{ad}} (t)$ for STIRAP with a constant gap. 

Let us consider the Magnus expansion of $\hV_\uI^{(2,\mm{opt})} (t) = \hV_\uI (t) + \hW_{1,\mm{ad}}
(t) + \hW_{2,\mm{opt},\mm{ad}} (t)$ up to third order [see Eq.~\eqref{eq:magnus_series}], with
$\hV_\uI (t)$ given by Eq.~\eqref{eq:VintMagnus} and $\hW_{1,\mm{ad}} (t)$ by Eq.~\eqref{eq:H1ad}.
For convenience we parametrize $\hW_{2,\mm{opt},\mm{ad}} (t)$ as a time-derivate,
$\hW_{2,\mm{opt},\mm{ad}} (t) = i \rd_t \hzeta (t)$. Since $\hW_{1,\mm{ad}} (t)$
[Eq.~\eqref{eq:H1ad}] cancels leakage at first order, we require $\hW_{2,\mm{opt},\mm{ad}} (t)$ to
cancel leakage at second and third order (formally at orders $\epsilon^2$ and $\epsilon^3$).  Hence,
the differential equation determining $\hW_{2,\mm{opt},\mm{ad}} (t)$ is given by $\rd_t
\hOmega_2^{(2,\mm{opt})} (t) + \rd_t \hOmega_3^{(2,\mm{opt})} (t) = \mathcal{O}(\epsilon^4)$. We
explicitly have
\begin{equation}
	\begin{aligned}
		&\rd_t \hzeta (t) + \rd_t \hOmega_2^{(1)} (t) +\frac{1}{2} 
		\left(\com{\rd_t \hOmega_1^{(1)}(t)}{\hzeta  (t)} + 
		\com{\rd_t \hzeta (t)}{\hOmega_1^{(1)} (t)} \right. \\
		&+ \left.  \com{\rd_t \hOmega_1^{(1)} (t)}{\hOmega_2^{(1)} (t)} \right) 
		-\frac{1}{6} \com{\hOmega_1^{(1)}(t)}{\rd_t \hOmega_2^{(1)}(t)}
		= \mathbf{0}.
	\end{aligned}
	\label{eq:opt2ndorder}
\end{equation}
Note that the above equation is not specific to STIRAP and holds whenever $\sq \hV (t) = \hV (t)$.
As explained in the main text, we could numerically solve Eq.~\eqref{eq:opt2ndorder} to find the
\textit{exact} $\hW_{2,\mm{opt},\mm{ad}} (t)$. However, since leaking out of the dark state
subspace is more harmful than a phase error, one can try to find an approximate solution to
Eq.~\eqref{eq:opt2ndorder} that only cancels \textit{pure} leakage at the expense of a phase
error.

We look for solutions of the form $\rd_t \hzeta = \gamma \rd_t \hOmega_2^{(1)}$. Substituting this
ansatz in Eq.~\eqref{eq:opt2ndorder} leads to 
\begin{equation}
	\begin{aligned}
	&(\gamma + 1) \rd_t \hOmega_2^{(1)} (t) + \frac{1}{2}(\gamma + 1) 
	\com{\rd_t \hOmega_1^{(1)}(t)}{\hOmega_2^{(1)} (t)} \\
	&+
	\frac{1}{2}\left(\gamma + \frac{1}{3}\right) 
	\com{\rd_t \hOmega_2^{(1)} (t)}{\hOmega_1^{(1)}(t)} = \mathbf{0}.
	\end{aligned}
	\label{eq:opt2ndorderlin}
\end{equation}
We notice that if we would choose $\gamma=-1$, we would find the same control as the one predicted
by Eq.~\eqref{eq:Hctrl2} and which corrects for phase errors.

To find a non-trivial choice for $\gamma$, we consider the integral of
Eq.~\eqref{eq:opt2ndorderlin}. We find
\begin{equation}
	(\gamma + 1) \hOmega_2^{(1)} (t) + \hOmega_{\mm{opt}} (t) = \mathbf{0},
	\label{eq:intopt2ndorderlin}
\end{equation}
where
\begin{equation}
	\begin{aligned}
	\hOmega_{\mm{opt}} (t) &= \int_0^t \ud t_1 \left\{\frac{1}{2}(\gamma + 1)
	\com{\rd_{t_1} \hOmega_1^{(1)}(t_1)}{\hOmega_2^{(1)} (t_1)} \right. \\
	&\phantom{={}}
	\left. + \frac{1}{2}\left(\gamma + \frac{1}{3}\right)
	\com{\rd_{t_1} \hOmega_2^{(1)} (t_1)}{\hOmega_1^{(1)}(t_1)}\right\}\\
	&= \left\{\frac{1}{2}\left[(\gamma+1) g_2 (t) + \left(\gamma+\frac{1}{3}\right) g_1 (t) \right]
	\ketbra{\ud}{\ub_-} \right. \\
	&\phantom{={}} + \frac{1}{2}\left[(\gamma+1) g_2^\ast (t) + \left(\gamma+\frac{1}{3}\right)
	g_1^\ast (t) \right] \ketbra{\ud}{\ub_+} \\
	&\phantom{={}} 
	\left.
	\vphantom{ + \frac{1}{2}\left[(\gamma+1) g_2^\ast (t) +
	\left(\gamma+\frac{1}{3}\right)g_1^\ast (t) \right]}
	+ \mm{H.c.}\right\}.
	\end{aligned}
	\label{eq:intsol}
\end{equation}
The functions $g_1 (t)$ and $g_2 (t)$ are given by
\begin{equation}
	g_1 (t) = -\frac{1}{2\sqrt{2}\Omega_0^2}\int_0^t \ud t_1 e^{i \Omega_0 t_1}
	\dot{\theta}^3 (t_1),
	\label{eq:g2}
\end{equation}
and
\begin{equation}
	\begin{aligned}
		g_2 (t) &= \frac{1}{2\sqrt{2} \Omega_0^2} e^{i \Omega_0 t}
		\dot{\theta} (t) \int_0^t \ud t_1
		\dot{\theta}^2 (t_1) \\
		&\phantom{=}- \frac{1}{2\sqrt{2} \Omega_0^2}\int_0^t
		\ud t_1 e^{i \Omega_0 t_1}
		\dot{\theta}^3 (t_1) \\
		&= g_0 (t) + g_1 (t).
	\end{aligned}
	\label{eq:intg1}
\end{equation}
Since $g_0 (t_\uf) = 0$, we can choose $\gamma=-2/3$ to cancel terms proportional to $g_1 (t)$ and
consequently suppress leakage at third order. With this choice, we have 
\begin{equation}
	\hOmega^{(2,\mm{opt})} (t_\uf) = \frac{1}{3}\hOmega_2^{(1)} (t_\uf) +
	\mathcal{O}(\epsilon^4),
	\label{eq:Omegarest}
\end{equation}
which shows that \textit{pure} leakage errors are suppress to third order at the expense of a phase
error.

\section{Approximate unitary dynamics for STIRAP with a time-dependent gap}
\label{sec:oscFG}

In this appendix, we derive an approximate expression for the infidelity of STIRAP using Gaussian
pulse shapes. Such an expression can be obtained using the Magnus expansion to find an approximate
time-evolution operator valid in the adiabatic regime ($\nu/G_0 < 1$).

\begin{figure}[t!]
	\includegraphics[width=\columnwidth]{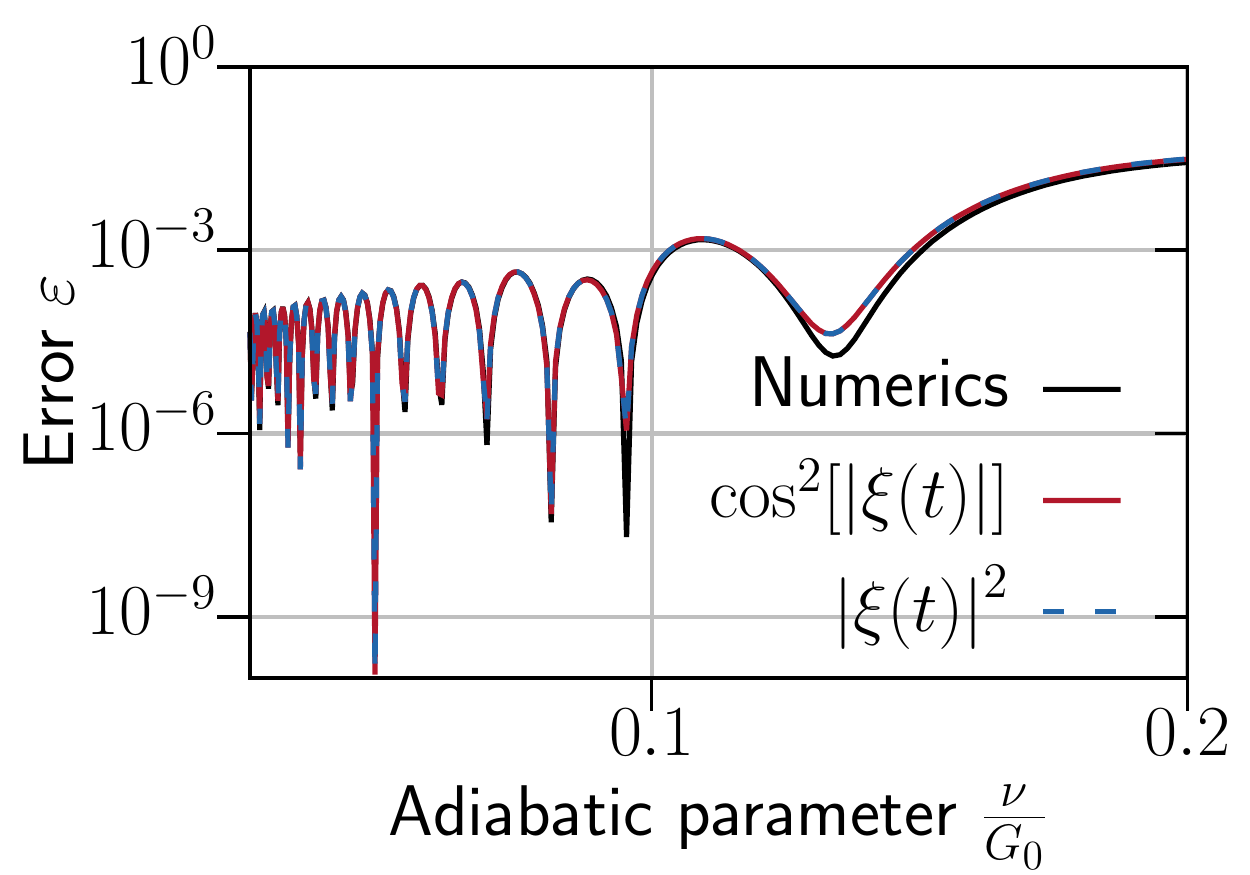}
	\caption{(Color online). Infidelity for STIRAP using Gaussian pulses. The approximate
		expression for the infidelity as given in Eq.~\eqref{eq:infidelityG} agrees with
		exact numerics in the adiabatic regime, $\nu/G_0 <1$.}
	\label{fig:compG} 
\end{figure}

We look for solutions of the time-dependent Schrödinger equation defined by
Eq.~\eqref{eq:HlambdaadG}. Using the general framework introduced in Eq.~\eqref{eq:Ugen}, we
factorize the evolution operator into $\hU (t) = \hU_0 (t) \hU_\uI (t)$, where 
\begin{equation}
	\begin{aligned}
	\hU_0 (t) &= \exp\left[-i \int_0^t \ud t_1 \hH_0 (t_1)\right] \\
	&= \ketbra{\ud}{\ud} + \exp[-i \Delta (t)] \ketbra{\ub_+}{\ub_+} + \exp[i \Delta
	(t)] \ketbra{\ub_-}{\ub_-},
	\end{aligned}
	\label{eq:U0Gaussian}
\end{equation}
with $\hH_0 (t)$ defined in Eq.~\eqref{eq:HlambdaadG}, and $\hU_\uI (t)$ is the evolution generated
by Eq.~\eqref{eq:VIG}. 

Using a first order Magnus expansion to approximate $\hU_\uI (t)$, we find 
\begin{equation}
	\hU_\uI (t) \simeq \exp\left[-i \int_0^t \ud t_1 \hV_\uI (t_1)\right].
	\label{eq:Magnus1G}
\end{equation}

Noticing that the transition probability from $\ket{1}$ to $\ket{3}$ is equivalent to the
probability to remain in the dark state $\ket{\ud}$ at $t_\uf$, we have 
\begin{equation}
	P_{\ket{1}\to \ket{3}} = P_{\ket{\ud} \to \ket{\ud}} 
	= \abs{\bra{\ud} \hU_\uI (t_\uf) \ket{\ud}}^2 = \cos^2 \left(\abs{\xi(t_\uf)}\right),
	\label{eq:P31G}
\end{equation}
where 
\begin{equation}
	\xi (t) = \int_0^t \ud t_1 \exp[i \Delta (t_1)] \dot{\theta} (t_1).
	\label{eq:argcosG}
\end{equation}

The state-dependent infidelity can then be expressed as 
\begin{equation}
	\varepsilon = 1 - P_{\ket{1} \to \ket{3}} \simeq \abs{\xi(t_\uf)}^2.
	\label{eq:infidelityG}
\end{equation}
Since $t_\uf$ is finite and $t_\uf \propto 1/\nu$, $\varepsilon$ exhibits an oscillatory behavior as
a function of $\nu/G_0$.

In Fig.~\ref{fig:compG}, we compare the infidelity as given by Eq.~\eqref{eq:infidelityG} with the
infidelity obtained through numerical integration of the Schrödinger equation defined by
Eq.~\eqref{eq:lambdasysH} with $G_\up (t)$ and $G_\us (t)$ given in Eq.~\eqref{eq:gaussian}.

\section{Matrix elements of \texorpdfstring{$\hV_{\mm{SAD}} (\tau)$}{VSAD}}
\label{sec:Vleaklzsadel}

The matrix elements of $\sq \hV_{\mm{SAD}} (\tau)$ for $\eta_{12} = \eta_{03} = \eta_{23} = \eta$
are given by

\begin{widetext}
\begin{equation}
	\begin{aligned}
		V_{12} (\tau) = \bra{\tilde{1}} \sq \hV_{\mm{SAD}} (\tau) \ket{2} 
		&= \frac{\eta ^2 \left(\tau -\sqrt{\eta ^2+\tau
		^2}\right)}{\sqrt{\left(\tau -\sqrt{\eta ^2+\tau ^2}\right)^2+\eta ^2} \sqrt{\left(2
		\left(\eta ^2+\tau ^2\right)^{3/2}+\sqrt{4 \left(\eta ^2+\tau
		^2\right)^3+\eta ^2}\right)^2+\eta ^2}}\\
		&\phantom{={}}
		+\frac{i \eta  \left(\sqrt{\eta
		^2+\tau ^2}+\tau \right) \left(2 \left(\eta ^2+\tau ^2\right)^{3/2}+\sqrt{4
		\left(\eta ^2+\tau ^2\right)^3+\eta ^2}\right)}{\sqrt{\left(\sqrt{\eta
		^2+\tau ^2}+\tau \right)^2+\eta ^2} \sqrt{\left(2 \left(\eta ^2+\tau
		^2\right)^{3/2}+\sqrt{4 \left(\eta ^2+\tau ^2\right)^3+\eta
		^2}\right)^2+\eta ^2}},
	\end{aligned}
	\label{eq:Vleaklzsad13}
\end{equation}
\begin{equation}
	\begin{aligned}
		V_{13} (\tau) = \bra{\tilde{1}} \sq \hV_{\mm{SAD}} (\tau) \ket{3} 
		&= -\frac{\eta ^3}{\sqrt{\left(\tau -\sqrt{\eta ^2+\tau
		^2}\right)^2+\eta ^2} \sqrt{\left(2 \left(\eta ^2+\tau ^2\right)^{3/2}+\sqrt{4
		\left(\eta ^2+\tau ^2\right)^3+\eta ^2}\right)^2+\eta ^2}}\\
		&\phantom{={}}
		-\frac{i \eta ^2 \left(2\left(\eta ^2+\tau ^2\right)^{3/2}+\sqrt{4 \left(\eta ^2+\tau
		^2\right)^3+\eta ^2}\right)}{\sqrt{\left(\sqrt{\eta ^2+\tau ^2}+\tau
		\right)^2+\eta ^2} \sqrt{\left(2 \left(\eta ^2+\tau ^2\right)^{3/2}+\sqrt{4
		\left(\eta ^2+\tau ^2\right)^3+\eta ^2}\right)^2+\eta ^2}},
	\end{aligned}
	\label{eq:Vleaklzsad14}
\end{equation}
\begin{equation}
	\begin{aligned}
		V_{02} (\tau) = \bra{\tilde{0}} \sq \hV_{\mm{SAD}} (\tau) \ket{2} 
		 &= -\frac{\eta ^2 \left(\tau -\sqrt{\eta ^2+\tau
		 ^2}\right)}{\sqrt{\left(\tau -\sqrt{\eta ^2+\tau ^2}\right)^2+\eta ^2}
		 \sqrt{\left(\sqrt{4 \left(\eta ^2+\tau ^2\right)^3+\eta ^2}-2 \left(\eta ^2+\tau
		 ^2\right)^{3/2}\right)^2+\eta ^2}}\\
		 &\phantom{={}}
		 +\frac{i \eta  \left(\sqrt{\eta ^2+\tau ^2}+\tau
		 \right) \left(\sqrt{4 \left(\eta ^2+\tau ^2\right)^3+\eta ^2}-2 \left(\eta ^2+\tau
		 ^2\right)^{3/2}\right)}{\sqrt{\left(\sqrt{\eta ^2+\tau ^2}+\tau \right)^2+\eta ^2}
		 \sqrt{\left(\sqrt{4 \left(\eta ^2+\tau ^2\right)^3+\eta ^2}-2 \left(\eta ^2+\tau
		 ^2\right)^{3/2}\right)^2+\eta ^2}},
	\end{aligned}
	\label{eq:Vleaklzsad23}
\end{equation}
and 
\begin{equation}
	\begin{aligned}
		V_{03} (\tau) = \bra{\tilde{0}} \sq \hV_{\mm{SAD}} (\tau) \ket{3} 
		&= \frac{\eta ^3}{\sqrt{\left(\tau -\sqrt{\eta ^2+\tau
		^2}\right)^2+\eta ^2} \sqrt{\left(\sqrt{4 \left(\eta ^2+\tau ^2\right)^3+\eta ^2}-2
		\left(\eta ^2+\tau ^2\right)^{3/2}\right)^2+\eta ^2}}\\
		&\phantom{={}}
		-\frac{i \eta ^2 \left(\sqrt{4
		\left(\eta ^2+\tau ^2\right)^3+\eta ^2}-2 \left(\eta ^2+\tau
		^2\right)^{3/2}\right)}{\sqrt{\left(\sqrt{\eta ^2+\tau ^2}+\tau \right)^2+\eta ^2}
		\sqrt{\left(\sqrt{4 \left(\eta ^2+\tau ^2\right)^3+\eta ^2}-2 \left(\eta ^2+\tau
		^2\right)^{3/2}\right)^2+\eta ^2}}.
	\end{aligned}
	\label{eq:Vleaklzsad24}
\end{equation}

\end{widetext}

\end{appendix}

\bibliographystyle{apsrev4-1}

\end{document}